# Terahertz Nonlinear Optics in Two-dimensional Semi-Hydrogenated SiB

Ali Ramazani[1], Farzaneh Shayeganfar[2,3], Anshuman Kumar[4], Nicholas X Fang[1*]

[1]Department of Mechanical Engineering, Massachusetts Institute of Technology, Cambridge, MA 02139, USA
[2]Department of Civil and Environmental Engineering, Rice University, Houston, TX 77005, USA
[3]Department of Physics and Energy Engineering, Amirkabir University of Technology
[4]Physics Department, Indian Institute of Technology Bombay, Mumbai 400076, India

**Abstract**

Phase transition and current of planar clusters with loops of Josephson $\pi$ junction is of great importance for application in the adiabatic quantum computation devices (AQC). Each $\pi$- ring possesses an orbital moment with a current. The orientation of these orbital moments is altered by self-interactions and controlled with a bias current. In the current research, we construct the phase diagram of semi-hydrogenated SiB (H-SiB) semiconductor using a two- dimensional (2D) Ising model with considering nearest-neighbor and diagonal interactions of spin planes on the on the hexagonal lattice. Our findings reveal that by decreasing temperature, two phase transitions of order and disorder occur and at low temperatures a glassy state appears, where this model maps to spin qubit. We theoretically study the magnetostriction effect by considering of laser electromagnetic field coupled to the spin planes of electric polarization of magnetic insulator for both unstrained and strained H-SiB monolayers. The dynamic of laser-spin coupling and spin current shows nonlinear optical effects and high-harmonic generation (HHG) under both terahertz (THz) and gigahertz (GHz) electromagnetic waves. The findings of this research open an avenue for experimentalists how to detect HHG in the topological insulators, Dirac systems, Mott insulators, and superconducting memories.

[*]Corresponding Author; Professor, Department of Mechanical Engineering, MIT, Office: 617-253-2247; Email: nicfang@mit.edu

# I. INTRODUCTION

The phase diagram of two dimensional (2D) materials is complicated and involves several phases with different types of ordering. Ising model can be used to understand the phase behavior of 2D structures taking into account the spin interactions [1,2]. Indeed, Ising model is a well-known model for ferromagnetism and phase transition, as this model determines the critical point and the interaction strength of different phases [3]. An anisotropic next-nearest neighbor Ising (ANNNI) model describes a cubic lattice of Ising spins and spin-spin interaction [1,4,5]. Arrays of π ring, which are combinations of high and low temperature superconducting materials deposited on the substrate, apply the Ising model with competing antiferromagnetic interactions applicable in the adiabatic quantum computing (AQC) [6,7]. A superconducting loop consisting of Josephson junctions is named as a single π-ring [6]. The odd number of π junctions in the loop results in a π shift and orbital current or magnetic current at the ring, which circulates in counterclockwise or clockwise direction [6]. A planar array of electrically isolated π-rings could be considered as Ising spins or magnetic moments perpendicular to the plane, whereas the dipole-dipole interaction between leads to a formation of the disordered structures in a 1D chain [8].

The planar clusters of π-rings such as square lattices have been used for AQC by applying competing interaction of Ising model [9]. These interactions depend on the coupling between π-rings, which are controllable boundary condition and topological defects. Several problems can be encoded into the ground state of the Ising model with distribution of coupling between their sites [9]. The adiabatic evolution of the ground state of the π-ring systems is taken into account by the AQC procedure by consideration of changing coupling constants, when the initial state evolves to final state [9].

The high-harmonic generation (HHG) as nonlinear optical phenomenon in semiconductors has been studied by several groups [10-12]. A strong mid-infrared laser field is recently available [13], where smaller photon energy of input laser compared to the band gap creates nonlinear dynamics and causes HHG generation [13-19]. Progress in solid state HHG studies in semiconductors involves Dirac systems, Mott insulators, topological insulators and charge-density-wave materials. Recent development of terahertz (THz) and gigahertz (GHz) lasers reveals that such low frequency lasers can create the magnetic excitations [20]. The THz and GHz lasers lead to the magnetic dynamics [21], which confirms that the THz/GHz photon energy can excite the magnetic states in magnetic insulators. Therefore, GHz and THz waves play a significant role in the generation of HHGs in semiconductors with spin interactions. For instance, a THz laser can generate the second harmonic generation (SHG) from magnetic excitations in antiferromagnetic insulator.

Among 2D materials, semi-hydrogenated hexagonal SiB (H-SiB) monolayer shows a semiconducting behavior and its tensile and compressive strained structures play a pivotal role in a wide range of optoelectronic devices including plasmonic sensors and quantum computing [22-25]. In the present paper, we firstly consider planar clusters of π-rings of hexagonal H-SiB, which is described by the Ising model. The exact solutions of four-site cluster or a plaquette gives the characteristic features of this model, whereas by mean-field approach we investigate the infinite hexagonal lattice of spin configuration and possible topological defects. In the next step, we move to calculate the phase diagram by applying Monte Carlo simulations for the 86 × 86 hexagonal lattices. Finally, we apply terahertz (THz) or gigahertz (GHz) electromagnetic waves to study HHG from spin current in the magnetic insulators.

**II. MODEL**

**A. Ising Model for Hexagonal unit cell**

We consider the following two-dimensional (2D) Ising model with the Hamiltonian taking into account two competing antiferromagnetic nearest-neighbor and next-nearest-neighbor interactions [26]:

$$H = J\sum_{<i,j>nn} s_i s_j + J'\sum_{<i,j>dn} s_i s_j - h \sum_i s_i \qquad (1)$$

where the constants **J, J'** > 0 are related to the elastic interaction on the H-SiB lattice as coupling constant of the nearest- and next-nearest-neighboring spins, $<i,j>nn$ and $<i,j>dn$ denote the summation over sites **i** and **j** being respectively nearest neighbors (**nn**) and diagonal neighbors (**dn**), and **h** is the analog of magnetic field corresponding to some external stress. The hexagonal H-SiB model is schematically illustrated in **Fig. 1**.

Such a model (**Eq. 1**) correctly describes a planar array of π-rings deposited onto an insulating substrate. At each π-ring, Ising spin creates a magnetic orbital moment. Such orbital moments are perpendicular to the plane and interact with each other via the long-range dipole-dipole interactions, which shows a minimum at $1/r^3$ (r is the distance between dipole moments) [26,28]. Therefore, we only consider interactions between the nearest- and next-nearest-neighboring spins by applying two constants of antiferromagnetic interactions, J and J'. In this study we determine the phase transition for different values of J/J', which range from 0 to 8. In fact, since the deformation energy of each atomic distortion decreases as $1/r^3$, the J/J' ratio for hexagonal structure should be equal to J/J′= $3^{3/2}$ [26].

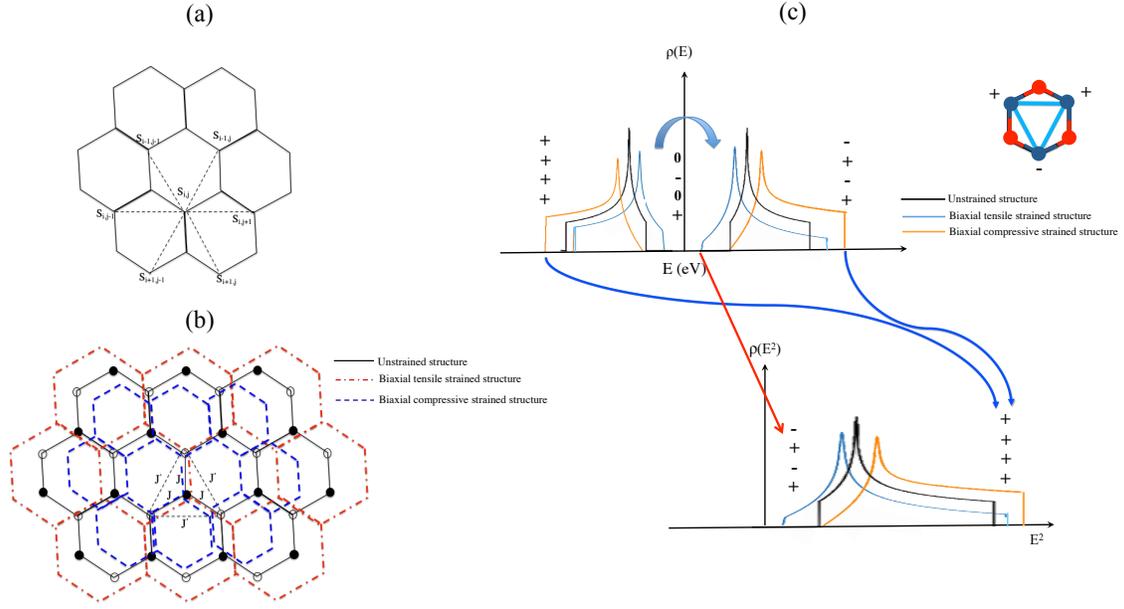

**Fig. 1:** (a) The geometry of Ising model with nearest neighbor J (solid lines) and diagonal J′ (dashed lines) interactions for the honeycomb lattice. (b) Unstrained (black) and tensile strained (red) structures of Ising model. (Fonts of the figure text needs to be bigger), and (c) Sketch of density of states ρ(E) of strained and unstrained hexagonal H-SiB lattice, which is transformed into triangular lattice ρ($E^2$), creating folding states by arrows around E = 0; ± signs represent the central states, band edges and phase differences among sites (adopted from Ref. [27]).

We consider some key parameters to analyze different phases in the strained H-SiB. The domain-wall length ($\mathcal{DL}$) defines as:

$$DL = \frac{1}{N} \sum_{<ij>_{nn}} (s_i s_j + |s_i s_j|) = \frac{1}{N} \sum_{<ij>_{nn}} (s_i s_j + 2) \qquad (2)$$

In **Eq. 2,** the sum is taken over the neighboring pairs, and N is a number of lattice sites. The $\mathcal{DL}$ for ferromagnetic phase, when all atoms of H, Si, B are displaced in one direction, becomes 4 and for antiferromagnetic phase is equal 0. The second key parameter for phase diagram is magnetization ($\mathcal{M}$), which is

$$M = |\frac{1}{N} \sum_{ij} s_{ij}| \tag{3}$$

and the last key parameter is the energy fluctuation or heat capacity peaks, which is explained in details in subsection B by **Eq. 10, and Eq. 11**.

### B. Thermal behaviors of four-site plaquette (Partition function, free energy, and peaks in specific heat)

Now let's discuss the four site plaquette at finite temperatures. If we know the set of energy levels and their degeneracy (**Fig. 2**), we can write the partition function in the form of (the Boltzmann constant is taken to be equal to one):

$$Z = 2(e^{(6J-3J')/T} + 3e^{(2J+J')/T} + 3e^{(-2J+J')/T} + e^{(-6J-3J')/T} +$$

$$e^{(6J-12J')/T} + 3e^{(+2J+4J')/T} + 3e^{(-2J+4J')/T} + e^{(-6J-12J')/T}) \tag{4}$$

The peaks in the heat capacity curves $C(T) = -T \frac{\partial^2 F}{\partial^2 T}$; (where F is the free energy (F = −T ln Z)) at different values of J/J' can be treated as manifestations of phase transitions, which are going to occur for the infinite lattice. From this viewpoint, we should analyze the low-temperature behavior of the heat capacity. The partition function becomes [9]:

$$Z = A' e^{A/T} (1 + Be^{-\frac{D}{T}} + \cdots) \tag{5}$$

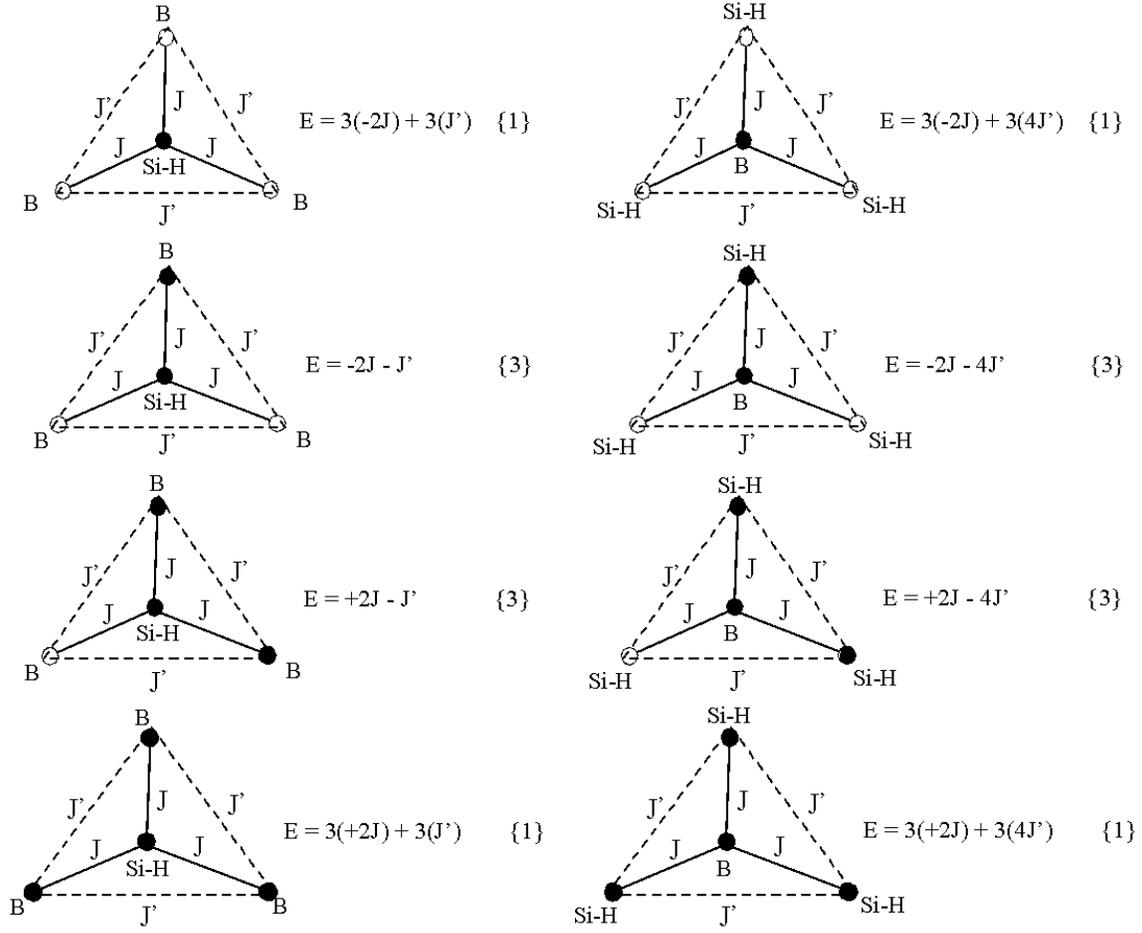

**Fig. 2.** Possible spin configurations of a four-site plaquette and the corresponding energies; the degree of degeneracy of each energy value is shown in the brackets. Only 16 configurations with the filled circle in the center are shown; there are also 16 similar configurations with the open circle in the center of the plaquette.

From **Eq. 3**, it is obvious that C(T) depends on neither *A* nor *A′*. Therefore, in order to identify Z at low-temperature zone (limit), we should only take into account the terms proportional to ($e^{-D/T}$). Heat capacity C(T) can be described as [9]:

$$C(T) \approx \frac{BD^2}{T^2} e^{-D/T} \tag{6}$$

In **Eq. 6** the terms ~ $e^{-D/T}/T$ are canceled for low temperatures. Using **Eq. 6** at

dC/dT= 0 condition, we can find temperature $T_c$, which is corresponding to the position of the heat capacity peak when T→0 [9]:

$$T_C = \frac{D}{2} \qquad (7)$$

For convenience, we introduce the dimensionless variables such as f = F/J'; $\alpha$ = J/J' and t = T/J' and rewrite the partition function in **Eq. 5** using these variables. The low temperature limit of the partition function separates into two regions for peak of heat capacity (for more details, please see **Ref. [9]**). At $\alpha = \frac{J}{J'} > 4$, we can write the partition function in the form similar to **Eq. 5**,

$$Z \approx 2\, e^{\frac{6J-3J'}{T}}(1 + 3e^{-\frac{D_1}{T}} + \cdots) \qquad (8)$$

where $D_1 = 4J - 16J'$. In the case of $\alpha = \frac{J}{J'} < 4$, the partition function becomes:

$$Z \approx 6\, e^{\frac{2J+4J'}{T}}(1 + \frac{1}{3}e^{-\frac{D_2}{T}} + \cdots) \qquad (9)$$

where $D_2 = 7J' - 4J$. As a result, using **Eq. 7**, we find the following positions of the heat capacity peaks in the low-temperature limit:

$$\alpha = \frac{t}{2} + \frac{16}{4}; \qquad for\ \alpha > 4 \qquad (10)$$

$$\alpha = \frac{-t}{2} + \frac{7}{4}; \qquad for\ \alpha < 4 \qquad (11)$$

**Eq. 10** indicates the α = J/J' as function of t = T/J' for $\alpha > 4$, which plotting this equation yields the positions of heat capacity maxima C(t, α) in the t-α plane at 16/4. **Eq. 11** similar to **Eq. 10** shows α(t) as a phase diagram of modeled plaquette of infinite systems for α < 4, where the cross over point will be at 7/4. To discover the order of phase transition in the H-SiB, we plotted the C(T) versus T/J' in **Fig. 3** for two limits of $\alpha$ as described in **Eq. 10 ($\alpha > 4$)**, and **Eq. 11 ($\alpha < 4$)**. As can be seen in this figure, C(T) has discontinuity for $\alpha < 2$ and $\alpha > 4$, confirming the second order transition due to the fact that C(T) is the second derivative of free energy.

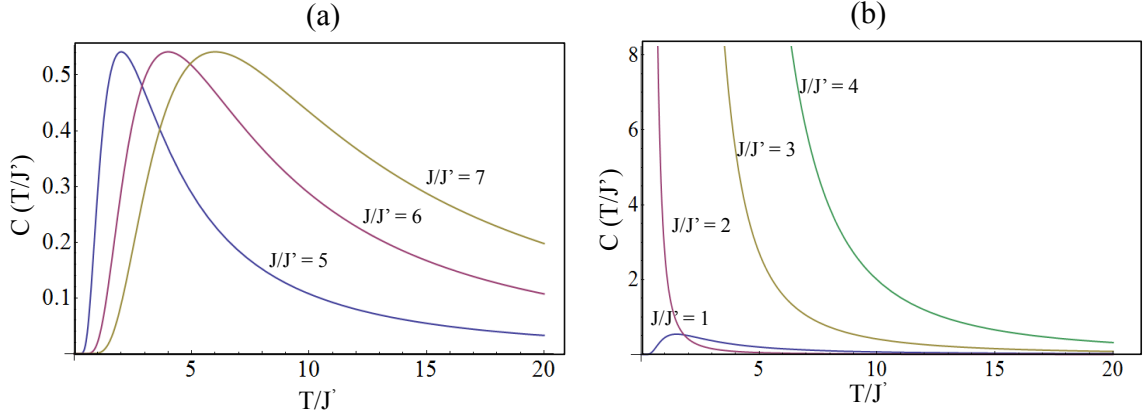

**Fig. 3:** (a) 2D plot of C(T) versus T/J' for $\alpha > 4$ (**Eq. 10**), and (b) for 2D plot of C(T) versus T/J' for $\alpha < 4$ (**Eq. 11**).

**C. Hexagonal infinite lattice analysis and energetics of topological defects:**

Now, we move to discuss the possible spin configurations of infinite hexagonal lattice model of semi-hydrogenated SiB corresponding to Hamiltonian **Eq. 1.** For this purpose, we model two configurations in **Fig. 4** at large and small ratio of coupling J/J', which have the lowest energy per site. The details of the model development to study the effect of topological defects including defected boundaries and dislocations on the lattice energy is reported in **Supporting Information (SI)**.

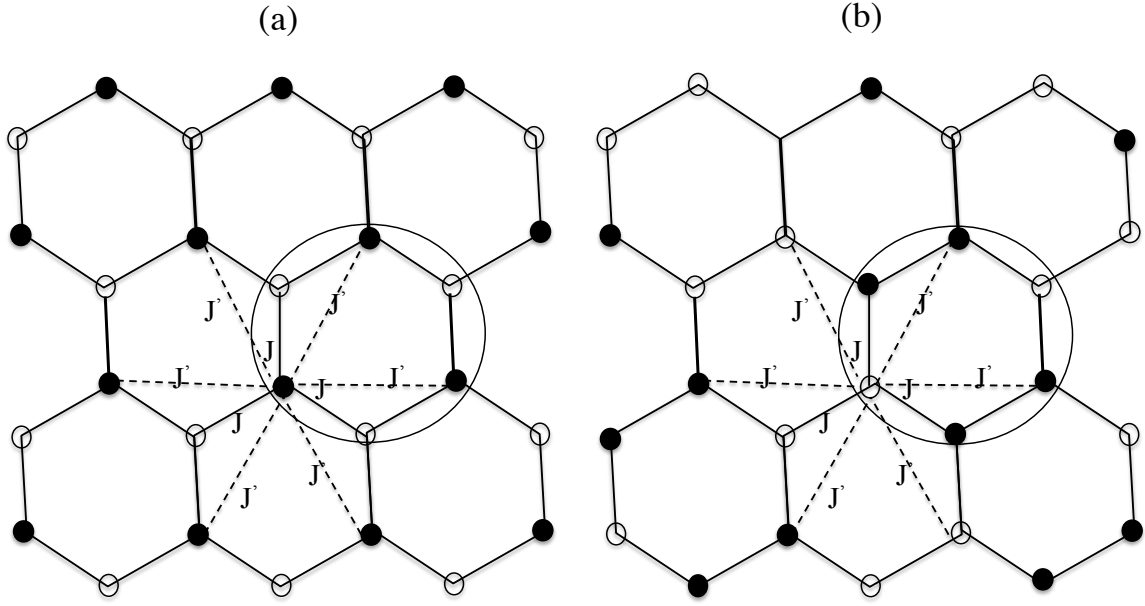

**Fig. 4:** (a) Single stripes oriented in three equivalent directions (antiferrodistorsive) that generate minimum energy for large **J > J'** values as acquired in **Eq. 12.** (b) Double stripes oriented along only one of the three possible directions (frustration) that generate minimum energy for **J < J'** as shown in **Eq. 12.**

Let's consider now a general case of this model, where different phases are arising at an arbitrary value of **J/J'**. When the interaction between the nearest neighbors dominates (**J >> J'**), the minimum energy corresponds to the antiferrodistorsive structure shown in **Fig. 4a**. The next-nearest-neighbor interaction leads to frustrations, and the possible structure favorable at small **J<<J'**, which is demonstrated in **Fig. 4b**. The crossover between these two structures occurs at **J/J' = 4**.

**Fig. 4(a)** $\quad E = -4J + 16J' \quad for\ (H-Si)\ site$

or $\quad E = -4J + 4J' \quad for\ (B)\ site$ (12)

**Fig. 4(b)** $\quad E = -4J' \quad for\ (H-Si)\ site$

or $\quad E = -2J - 6J' \quad for\ (B)\ site$

### D. Adiabatic quantum computation

The coupling interaction J and J' and their ratio can be controlled in π–rings system by external agent[10]. For instance, by introducing additional current as plotted in **Fig. 5,** current loops between π–rings can change and control this coupling interaction. In this figure, we consider that the two opposite currents are flowing along the straight lines, which induce the magnetic field. This magnetic field influences the interaction of π–rings and the values J/J'. A primary candidate for flux qubit is the π-ring system with low noise and long coherence time [9].

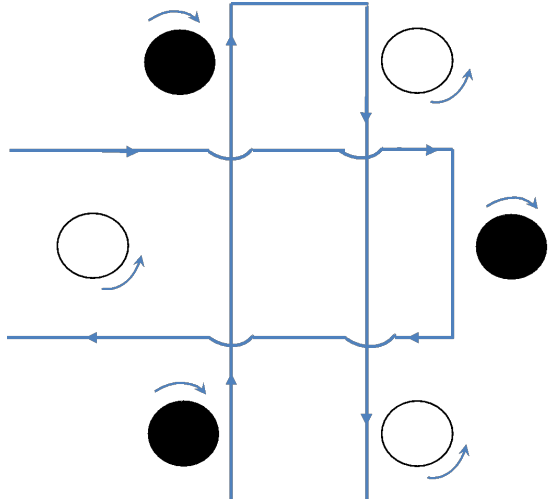

**Fig. 5:** The smallest hexagonal cluster of π-rings, 3×3. Filled and open circles denote clockwise and counterclockwise directed orbital currents associated with the orbital moments equal to s= +1 and s=−1, respectively. Here the value and the ratio of nearest-neighbor J and diagonal J' interactions can be controlled by the external bias current I. Arrows indicate the directions of currents in the π-rings and in control loops.

While the flexibility of this tunable coupling is presented in **Ref. 24,** the adiabatic quantum computation (AQC) [30], and the adjustable coupling of multi π-ring systems has been considered experimentally in **Ref. 25.** In line with experimental AQC studies [31], here we investigate the phase diagram of hexagonal π-ring system and explore different order and disorder states at different values of J/J' states, which can be tuned by controlling coupling.

### III. Phase Diagram

#### A. Monte Carlo simulation

Now, we consider a realistic and large system such as a 86×86 hexagonal lattice with Ising spins, where there are competing antiferromagnetic interactions at arbitrary value of $\alpha = J/J'$. We note that the physical phenomenon of the proliferation of defects can exist. We performed the numerical simulation of this system using the Monte Carlo method, which is the only method available to treat such large systems. We demonstrate the results of the Monte Carlo simulation at various temperatures $T/J' = 0.1–6.0$. The spin structure for the 86×86 hexagonal Ising lattice with competing antiferromagnetic interactions at $\alpha = J/J' = 2$ is shown in **Fig. 6**, where each segment in this figure corresponds to a π-ring. The dark segment corresponds to the up-spin orientation of the orbital moment; while the light segment corresponds to the down-spin orientation of the orbital moment, percolating to form very small islands or domain walls (**Fig. 8b ($\alpha = J/J' = 2$))**. At this value of $\alpha$, we see a complicated array, a mixture of light and dark segments. We guess that the two sublattice portions are mixed with the topological defects such as dislocations, which are created by shifting atoms by one lattice constant (**Fig. S2**), and domain boundaries.

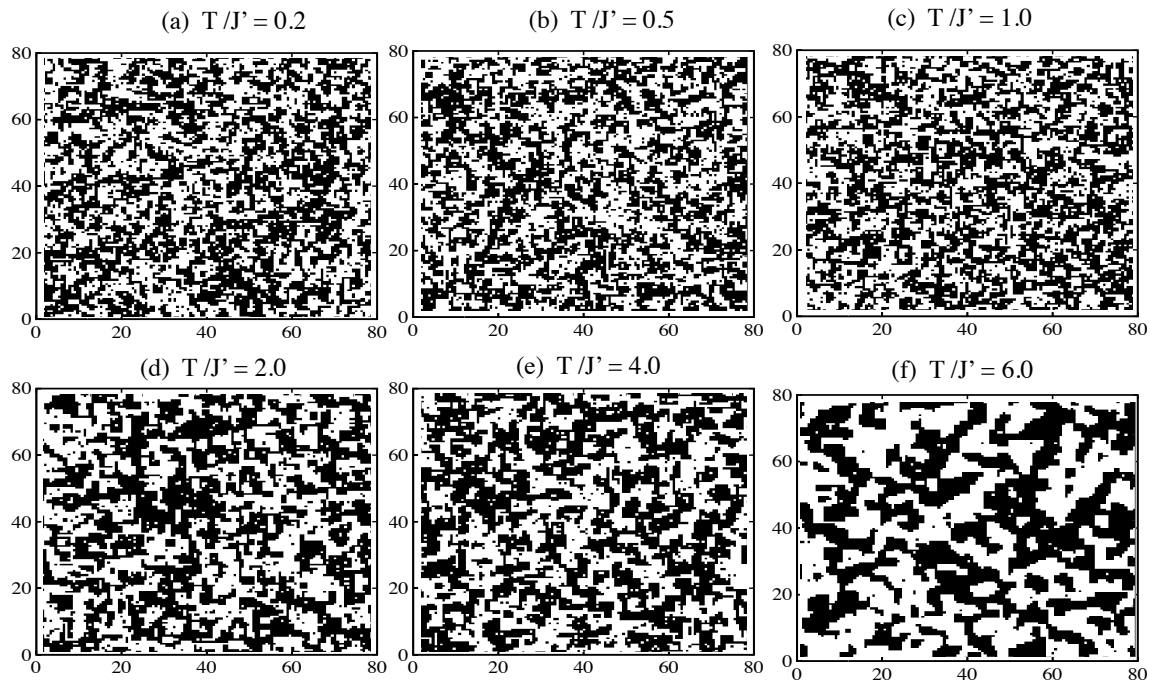

**Fig. 6:** Snapshots of the superstructures formed on a semi-hydrogenated SiB honeycomb lattice with J/J' = 2 at various T/J'.

The physical phenomenon of these complicated maps of **Fig. 6** may suggest a new type of disordered glassy state. Now let us investigate the process of evolution of 86 ×86 π-ring array when the coupling between all π-rings is tuned simultaneously in a way that the ratio α = J/J' = 5, which is the ratio of exchange constants in the corresponding Ising model, increases from zero (**Fig. 7**). First, we consider the value of T/J' is small, we have a well-defined ground state the stripe phase. The distribution of spins for such a state form stripes oriented horizontally or vertically.

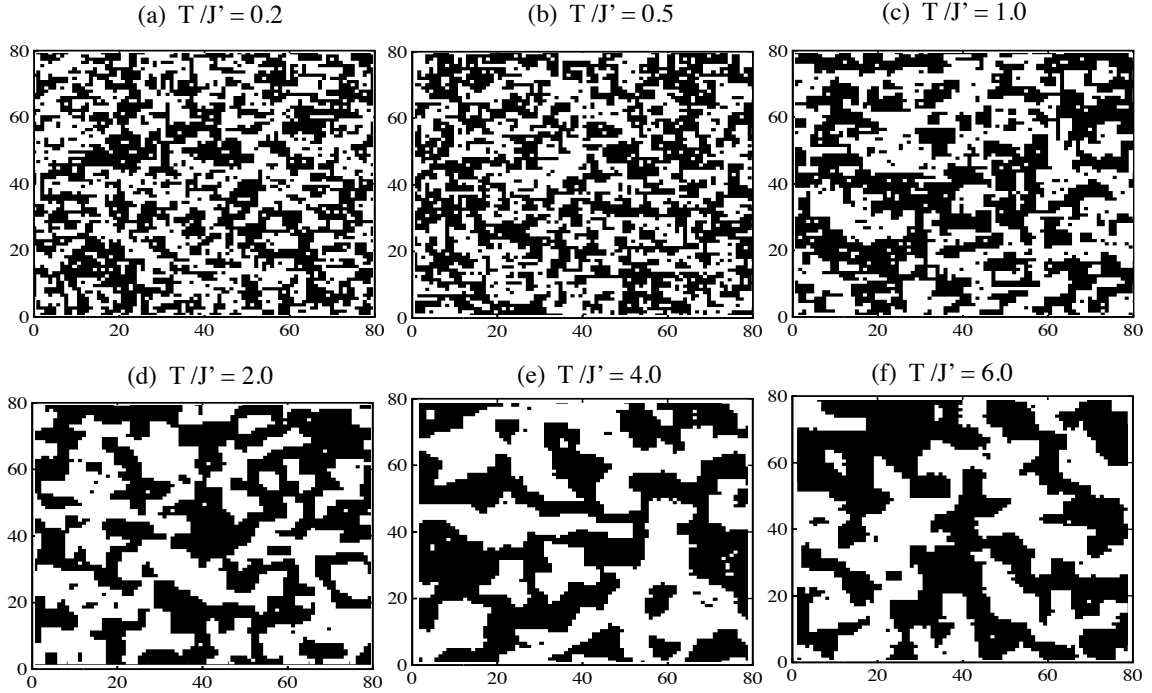

**Fig. 7:** Snapshots of the superstructures formed on a semi-hydrogenated SiB honeycomb lattice with J/J' = 5 at various T/J'.

Here, only in the vicinity of 0.2 < T/J' < 2, the system exhibits a pronounced disorder. The evolution of the physical system for 2 ≤ T/J' < 8 shows order states as shown in **Fig. 8**. These calculations indicate that at high temperatures, there will be less disorder than at low temperatures.

### B. Domain-wall Length ($\mathcal{DL}$)

The $\mathcal{DL}$ for the antiferromagnetically ordered state of the H-SiB (**Eq. 2**) becomes zero. For the ferromagnetically ordered state of H-SiB, when all spins of H, Si and B are displaced in one direction, the $\mathcal{DL}$ becomes 6. According to **Eq. 12**, for the J/J' << 4, the dominant state will be double stripe oriented (shown in **Fig. 4b**), which contributes to the minimum energy (E = -4J' for H-Si site and E = -2J-6J' for B site). The order parameter $\mathcal{DL}$ in this case becomes 2. Then, for $0 < \mathcal{DL} < 2$ and $2 < \mathcal{DL} < 6$ a disorder state will arise. The snapshots in **Fig. 6** and **Fig. 7** correspond to the $\mathcal{DL}>0$, i.e. for the disordered glassy state. A large number of such disordered configurations exist, where these configurations are trapped in the local minimum. All these local minima at very low temperature are separated by high-level barriers, which confirm that the disordered states are locally stable. The height of barriers decreases with increasing temperature, which is also confirmed by **Fig. 7**, where the order states in material increases with increasing temperature.

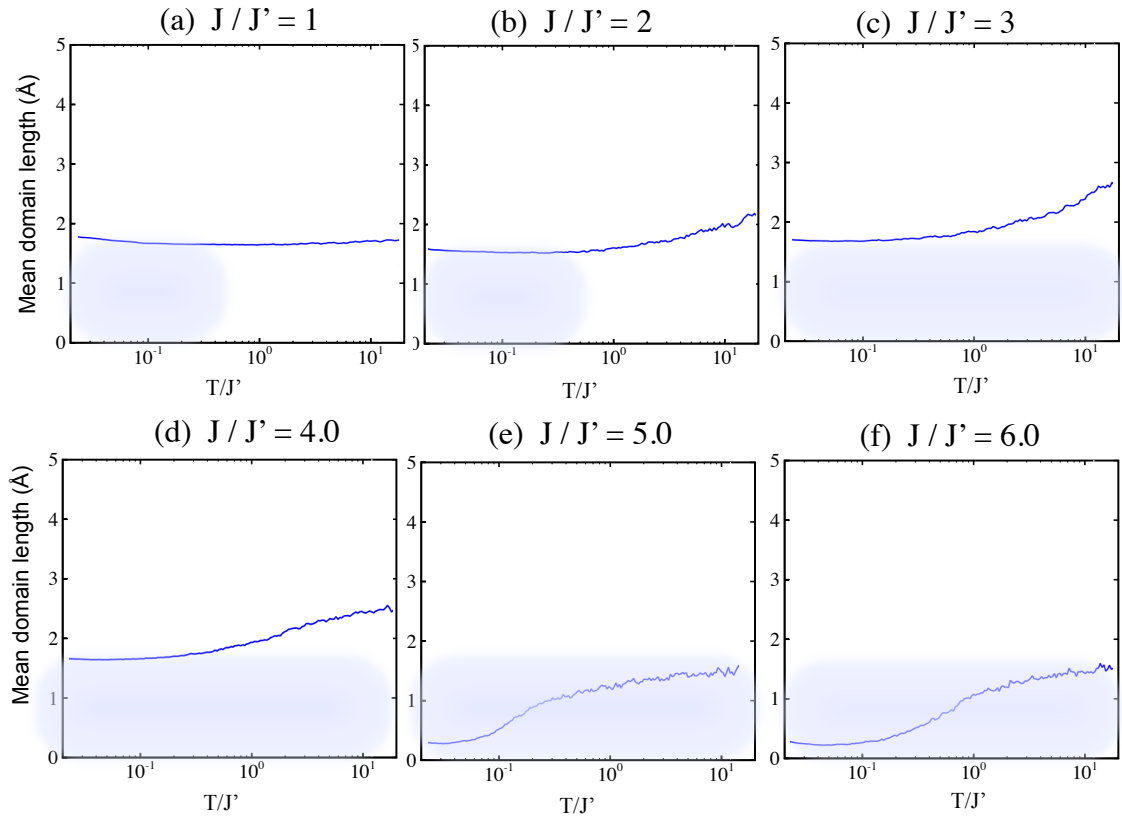

**Fig. 8.** The order parameter or mean length of domain-wall ($\mathcal{DL}$) as a function of temperature of 2D Ising model at different values of J/J' for the unstrained structure. Shadow regions show the transition temperature for glassy state.

We plotted **Fig. 8** for the order parameter or mean length of $\mathcal{DL}$ as a function of temperature at different values of J/J'. The steep deviation of $\mathcal{DL}$ of the honeycomb of H-SiB changes from 2 for J/J' ≤ 4 and from 0 for J/J' > 4. This result ($0<\mathcal{DL}<2$) is the signature of the disordered glassy state as plotted in figure 9. In **Fig. 9**, the phase diagram in **J/J' − T/J'** plane for H-SiB is depicted. This diagram is calculated using the Monte Carlo simulations. As can be seen in this figure, the phase diagram for small values of **J/J'** are consistent with the presented Monto Carlo simulation results for α = J/J'= 2 and α = J/J'= 5 in **Fig. 6** and **Fig. 7**, respectively.

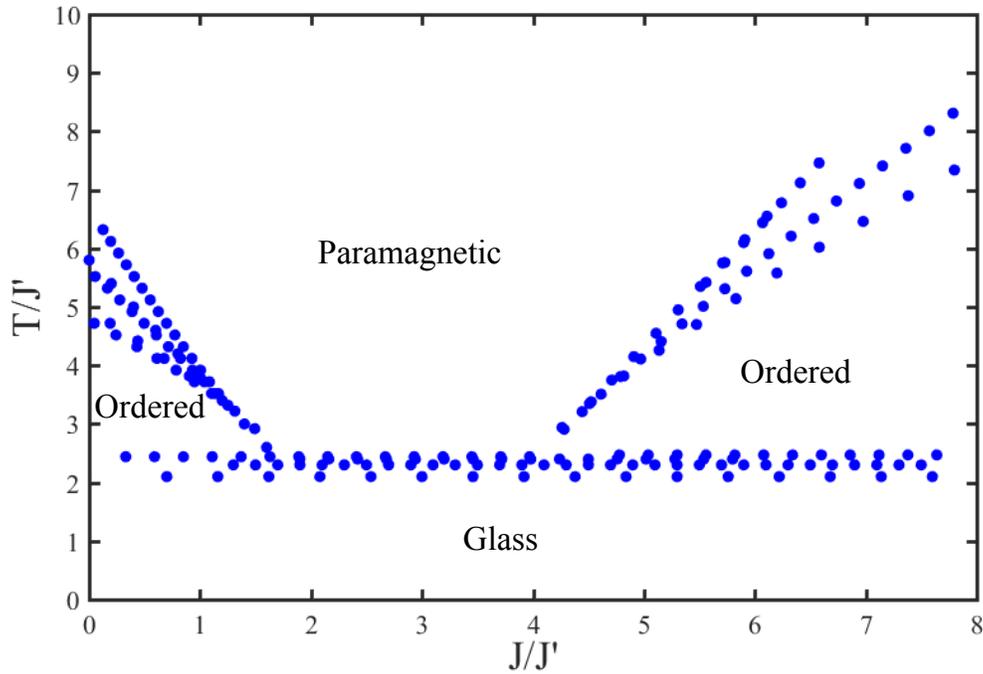

**Fig. 9:** Phase diagram for the semi-hydrogenated SiB model on a honeycomb lattice. The phase transition is calculated from the peaks in the specific heat and the change in the magnetization and domain-wall order parameter ($\mathcal{DL}$).

### C. Hamiltonian of strained hexagonal structures:

In this section, we extend the 2D Ising model for strained hexagonal H-SiB, in the hope that this model will open a new route to configure more complex phase behaviors of 2D nanostructures. This extension is in the direction of applying intramolecular interaction of spin-lattice of strained H-SiB. The extended Hamiltonian with applying spin-lattice term for strained hexagonal H-SiB defines as:

$$H = J\sum_{<i,j>nn} s_i s_j + J'\sum_{<i,j>dn} s_i s_j \pm f \sum_i s_i u_{ij} - h\sum_i s_i \qquad (13)$$

where **i** and **j** indexes are H-Si and B atoms, **J** and **J'** are the nearest-neighbor and next-nearest-neighbor coupling term of interaction between spins or the internal degrees of freedom. The "$f\, u_{ij}s_i$" is the spin-lattice coupling term [17], which can arise from the interaction forces of spins and displacements of atoms, which push them along the vertical direction, and h is related to the external field. In this equation, $u_{ij}$ is the elastic tensor has the form of:

$$u_{ij} = \frac{1}{2}(\partial_i u_j + \partial_j u_i + \partial_i h_a\, \partial_j h_a) \qquad (14)$$

and $h_a$ is out-of-plane displacements (deformation). We assume that $f$ decays exponentially as $f = f_0\, e^{-r/\xi}$, where $\xi$ is the length scale. Therefore, it's reasonable to take the nearest neighbor spin-lattice interaction in our simulation model. In the next step, we investigate the phase diagram by taking into account the spin-lattice interaction of **Eq. 13** and perform Monte Carlo simulation for analyzing different phase behaviors.

### D. Peaks in specific heat of strained Hamiltonian

We move to calculate the partition function, free energy, and peaks in specific heat in the low temperature limit by considering extended Hamiltonian (**Eq. 13**) for strained H-SiB hexagonal lattice. We write the partition function of possible spin configurations of a four-site plaquette by considering spin-lattice as plotted in **Fig. S3** Similar to **Fig. 2** in the form of (the Boltzmann constant is taken to be equal to one):

$$Z = 2\left(e^{\frac{(6J-3J'+3u)}{T}} + 3e^{\frac{(2J+J'+5u)}{T}} + 3e^{\frac{(-2J+J'+7u)}{T}} + e^{\frac{(-6J-3J'+9u)}{T}} + e^{\frac{(6J-12J'-3u)}{T}} + 3e^{\frac{(+2J+4J'+u)}{T}} + 3e^{\frac{(-2J+4J'+5u)}{T}} + e^{\frac{(-6J-12J'+9u)}{T}}\right) \quad (15)$$

As it has already been calculated (subsection B.), the heat capacity peaks in the low-temperature limit behaves as described in **Eq. 10**, and **Eq. 11**. By repeating the same calculation for strained Hamiltonian we obtain:

$$\alpha = \frac{t}{2} + 4 + \frac{4}{4}u; \qquad for\ \alpha > 4 \quad (16)$$

$$\alpha = \frac{-t}{2} + \frac{7}{4} - \frac{2}{4}u; \qquad for\ \alpha < 4 \quad (17)$$

where ***u*** = ***u**$_{ij}$*/***J'*** and ***u**$_{ij}$* is the strain tensor, which is equal to (+2%, +4%, +6%) for tensile strain and (-2%, -4%, -6%) for compressive strain.

### C. Phase diagram for strained H-SiB

We perform several simulation tests and plot the strained structure phase diagram. In our simulation, we took two different regimes related to **J** >> **J'** and **J** << **J'**. In the first regime, when *J'* is much smaller than *J*, and is around 0.01, the shift in the position of heat capacity as discussed earlier in **Eq. 16**, and **Eq. 17** becomes $\alpha = \frac{t}{2} + 4 + \frac{u_{ij}}{0.01}$ and $\alpha = \frac{-t}{2} + \frac{7}{4} - \frac{1}{2}\frac{u_{ij}}{0.01}$, respectively. As we can see in **Fig. 10** by taking j' = 0.01 and u$_{ij}$ = 0.02, there is a shift for phase region boundary. For instance, for J/J'>4 the boundary is shifted towards 6 and for J/J'<4, the boundary is shifted towards 0.5.

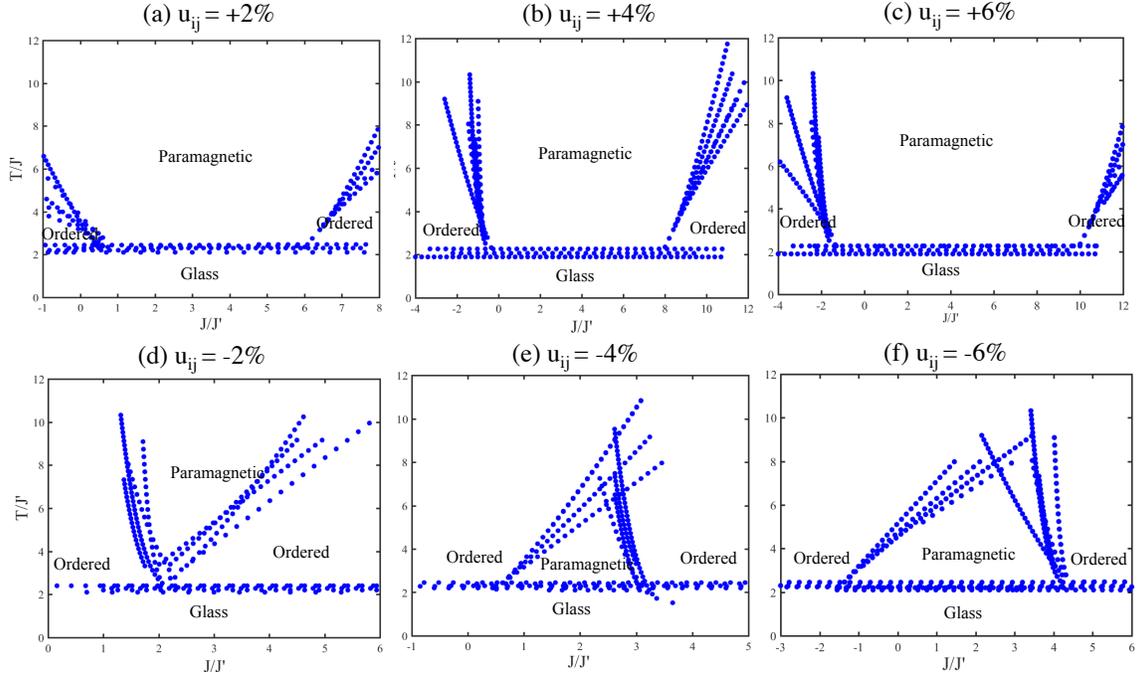

**Fig. 10:** Phase diagram for the strained semi-hydrogenated SiB model on a honeycomb lattice. The phase transition is calculated for J' = 0.01 and (a) $u_{ij}$ = +0.02, (b) $u_{ij}$ = +0.04, (c) $u_{ij}$ = +0.06, (d) $u_{ij}$ = -0.02, (e) $u_{ij}$ = -0.04 and (f) $u_{ij}$ = -0.06 from the peaks in the specific heat (**Eq. 16**, and **Eq. 17**) and the change in the magnetization and domain-wall order parameter ($\mathcal{DL}$).

To compare phase transition of unstrained and strained H-SiB structure, we compute and plot the phase transition for tensile strained structure with ($u_{ij}$ = +0.02, +0.04, +0.06) and for compressive strained structure with ($u_{ij}$ = -0.02, -0.04, -0.06) in **Fig. 10**. For instance, for 4% tensile strained structure by considering (J' = 0.01 and $u_{ij}$ = 0.04), the shift in the position of heat capacity as discussed earlier in **Eq. 16, and Eq. 17** becomes $\alpha = \frac{t}{2} + 4 + \frac{0.04}{0.01}$ and $\alpha = \frac{-t}{2} + \frac{7}{4} - \frac{1}{2}\frac{0.04}{0.01}$, respectively (**Fig. 10b**). Similarly for 6% tensile strained structure (J' = 0.01 and $u_{ij}$ = 0.06), the shift in the position of heat capacity becomes $\alpha = \frac{t}{2} + 4 + \frac{0.06}{0.01}$ and $\alpha = \frac{-t}{2} + \frac{7}{4} - \frac{1}{2}\frac{0.06}{0.01}$ (**Fig. 10c**) and for -2% compressive strained, the shift becomes $\alpha = \frac{t}{2} + 4 - \frac{0.02}{0.01}$ and $\alpha = \frac{-t}{2} + \frac{7}{4} + \frac{1}{2}\frac{0.02}{0.01}$ (**Fig. 10d**), for -4% compressive strained, the shift becomes $\alpha = \frac{t}{2} + 4 - \frac{0.04}{0.01}$ and $\alpha = \frac{-t}{2} + \frac{7}{4} + \frac{1}{2}\frac{0.04}{0.01}$ (**Fig. 10e**) and for -6% compressive strained, the shift becomes $\alpha = \frac{t}{2} + 4 - \frac{0.06}{0.01}$ and $\alpha = \frac{-t}{2} + \frac{7}{4} + \frac{1}{2}\frac{0.06}{0.01}$, (**Fig. 10f**). Moreover, in Fig. 11, the heat capacity C(T/J') is depicted as function of T/J' for different α's using **Eq. 16 for α>4** and **Eq. 17** for α<4 at 6% tensile and 6% compressive strains.

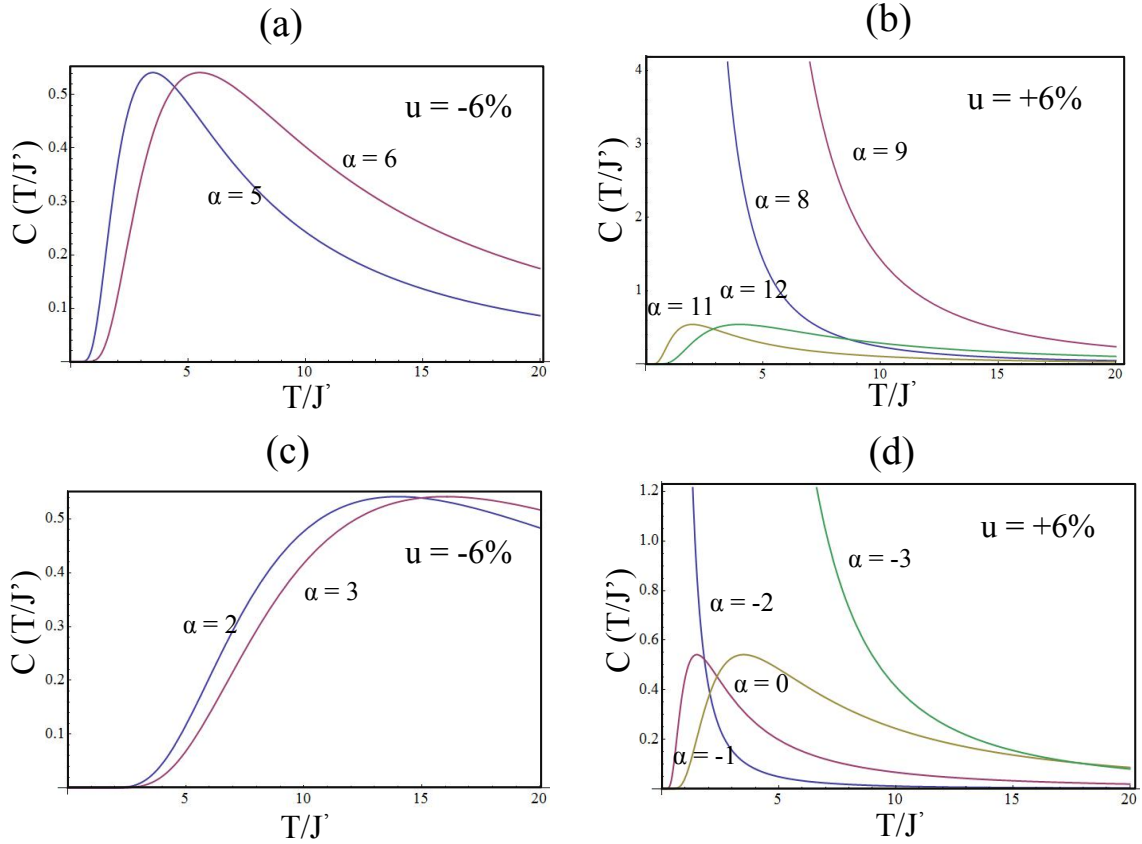

**Fig. 11:** Two-dimensional (2D) plot of heat capacity C(T/J') as a function of T/J' for (a) $\alpha > 4$ via **Eq. 16** at 6% compressive strain; (b) $\alpha > 4$ via **Eq. 16** at 6% tensile strain; (c) $\alpha < 4$ via **Eq. 17** at 6% compressive strain; and (d) $\alpha < 4$ via **Eq. 17** at 6% tensile strain.

In our previous study on the optoelectronic properties of H-SiB [22], we showed that for the strained structures, displacement of atoms and polarization are correlated to each other. For the tensile strain, by increasing the amount of applied strain, the gap energy and polarization both increases, while for the compressive strain, the gap energy and polarization both decreases with increasing the applied strain. In **Fig. 10(d,e,f),** paramagnetic (random orientation of Ising spins) state for compressive strains, which is considered as high symmetry state allowed two up and down states, decreases with increasing temperature. It is due to symmetry broken at high temperature for compressive strains and H-SiB experiences phase transition to ordered states. Moreover, in the compressive strained structures possess, the polarization is less than the tensile strained structures [22], which is in agreement with the reduction of bandgap under compressive strains, and the domination of electron-phonon interaction at high temperatures.

## IV. SPIN CURRENT AND SPIN TRANSFER TORQUE

Following our previous work on unstrained and strained semi-hydrogenated SiB, our findings reveal that application of strain on H-SiB result in a net dipole moment and induce polarization[1]. In fact, in strained structure, displacement of atoms causes dipole moment, and there is a correlation between polarization and displacement. In this study, polarized strained structure (H-SiB) coupled to spin creates spin current. The polarization operator ($\hat{P}$) defines as [20]:

$$\hat{P} = -e \left[ \sum_{<i,j>nn} s_i s_j + \sum_{<i,j>dn} s_i s_j \right] \tag{18}$$

where **'e'** is intrinsic electric field. The spin current as one of the observable interest depends on the coupling matrix of intrinsic filed with spins, which defines as:

$$\hat{H}_{coupl} = -e \left[ J \sum_{<i,j>nn} s_i s_j + J' \sum_{<i,j>dn} s_i s_j \right] \tag{19}$$

Then $\hat{I}_{spin}$ by considering the total Hamiltonian with strain term (**Eq. 13**) becomes:

$$\hat{I}_{spin} = \left[ (J-e) \sum_{<i,j>nn} s_i s_j + (J'-e) \sum_{<i,j>dn} s_i s_j \pm f \sum_i s_i u_{ij} \right] \tag{20}$$

### A. Fermionization

Our spin model can be mapped to noninteracting spin-less fermions by means of the Jordan-Wigner transformation [32]:

$$\hat{S}_j^+ = \prod_{i(<j)} (1 - 2\hat{c}_i^+ \hat{c}_i) \hat{c}_j \tag{21}$$

$$\hat{S}_j^- = \prod_{i(<j)} (1 - 2\hat{c}_i^+ \hat{c}_i) \hat{c}_j^+ \tag{22}$$

$$\hat{S}_j^z = \frac{1}{2} - \hat{c}_j^+ \hat{c}_j \tag{23}$$

$$\hat{S}_j^\pm = (\hat{S}_j^x \pm i\hat{S}_j^y)/2 \tag{24}$$

then we can rewrite the Ising Hamiltonian of **Eq. 1** as:

$$H = J \sum_{<i,j>nn} \sigma_i^z \sigma_j^z + J' \sum_{<i,j>dn} \sigma_i^z \sigma_j^z - h \sum_i \sigma_i^x \pm f \sum_i \sigma_i^x u_{ij} \tag{25}$$

$\sigma_i^z, \sigma_i^y, \sigma_i^x$ are Pauli matrices. In our study, there is no external electric field, then the last term will be zero ($h \sum_i \sigma_i^x = 0$) and we suppose that J' = 0 and $F = \pm f u_{ij}$. The (anti-) commutation relation is:

$$\{c_i, c_j^+\} = \delta_{ij}, \{c_i, c_j\} = \{c_i^+, c_j^+\} = 0 \tag{26}$$

The following step is to change the spin axes, so that $\hat{\sigma}_i^z \to -\hat{\sigma}_i^x$ and $\hat{\sigma}_i^x \to \hat{\sigma}_i^z$. This is primarily to simplify the algebra in future calculations. Note that this rotation does not influence the physics, although it changes the appearance of the Hamiltonian, and the interpretation of the several eigenvalues of the Pauli matrices. This rotation of the spin axis is equivalent with substituting the following expressions for $\hat{\sigma}_i^z$ and $\hat{\sigma}_i^x$ in the Hamiltonian **Eq. 26** as it is written in:

$$\hat{\sigma}_j^z = \frac{1}{2} - \hat{c}_j^+ \hat{c}_j \tag{27}$$

$$\hat{\sigma}_j^x = \prod_{i(<j)} (1 - 2\hat{c}_i^+ \hat{c}_i)(\hat{c}_j + \hat{c}_j^+) \tag{28}$$

$$H = -J \sum_i \quad (\hat{c}_i^+ \hat{c}_{i+1} + \hat{c}_{i+1}^+ \hat{c}_i + \hat{c}_i^+ \hat{c}_{i+1}^+ + \hat{c}_{i+1} \hat{c}_i - 2F \hat{c}_i^+ \hat{c}_i + F) \qquad (29)$$

To gain a better insight to the Ising Hamiltonian of **Eq. 26,** we introduce the Fourier transformations and Bogoliubov transforming in SI section S3 to get exact solution, where the Hamiltonian turns out to be diagonalized by this choice of basis:

$$H_I = \sum_K \quad \epsilon_k \left( \gamma_k^+ \gamma_k - \frac{1}{2} \right) \qquad (30)$$

$$\epsilon_K = 2J \sqrt{1 + F^2 - 2F \cos(ka)} \qquad (31)$$

The spin current as defined in **Eq. 20** depends on the coupling term, and its matrix representation is:

$$H = 2J \sum_k \quad (F - \cos(ka)) \sigma_z + \sin(ka) \sigma_y \qquad (32)$$

$$\hat{I}_{spin} = (J - e) \sum_k \quad (F - \sin(ka)) \sigma_z + \cos(ka) \sigma_y \qquad (33)$$

In the case of unstrained structure, we set the F = 0; and the spin current will be:

$$\hat{I}_{spin} = (J - e) \sum_k \quad -\sin(ka) \sigma_z + \cos(ka) \sigma_y \qquad (34)$$

Now, we consider to spin current for the Hamiltonian **Eq. 20** with taking into account J'. In this case, the Hamiltonian of **Eq. S6 (SI)** becomes:

$$H =$$

$$\sum_k \quad 2(JF - J\cos(ka)) \hat{c}_k^+ \hat{c}_k + J\sin(ka) (\hat{c}_{-k}^+ \hat{c}_k^+ - \hat{c}_{-k} \hat{c}_k) +$$

$$2(J' \cos(2ka)) \hat{c}_k^+ \hat{c}_k + J' \sin(2ka) (\hat{c}_{-k}^+ \hat{c}_k^+ - \hat{c}_{-k} \hat{c}_k) - JF$$

(35)

after lengthy calculation, the Hamiltonian and spin current becomes:

$$H =$$

$$\sum_k \quad 2J \left[ (F - \cos(ka) \sigma_z + \sin(ka) \sigma_y \right] + 2J' \left[ -\cos(2ka) \sigma_z + \sin(2ka) \sigma_y \right]$$

(36)

$$\hat{I}_{spin} = \sum_k \quad (J-e)\left[(F-\sin(ka)\,\sigma_z + \cos(ka)\,\sigma_y\right] + (J')\left[(-\sin(2ka)\,\sigma_z + \cos(2ka)\,\sigma_y\right] \qquad (37)$$

## V. HIGH-HARMONIC GENERATION BY ELECTRIC POLARIZATION

In this section, we compute the high-harmonic generation (HHG) as a nonlinear optical effect from spin current in magnetic insulators under terahertz (THz) or gigahertz (GHz) electromagnetic waves. The coupling between spins and the electric polarization in the Ising model of H-SiB leads to study the dynamics of the spin array exposed to an external ac electric or magnetic field. Quantum master equation gives the time evolution of the spin current, which includes clear harmonic peaks.

From the experimental viewpoint, one key parameter is the strength of THz or GHz electromagnetic waves for HHG. Acquiring this strength help us to control and tune other nonlinear optical phenomena such as spin current rectification [15,33,34] and Floquet engineering of magnets [35-38].

To describe **the laser-spin coupling,** the first is the Zeeman coupling of the laser magnetic field **B(t)** along the $S^z$ direction, which reads as [20]:

$$\hat{H}^z_{ext}(t) = -B(t)\sum_j \quad [\eta_z^u + (-1)^j \eta_z^s]\hat{S}_j^z \qquad (38)$$

where the $\eta_z^u = g\mu_B$ ; ($\mu_B$ is the Bohr magneton and g is a gyroscope factor) and $\eta_z^s \neq 0$ acts as inner magnetic field on site j, we set ℏ =1 in this study. We can simplify the **Eq. 38** as [20]:

$$\hat{H}^z_{ext}(t) = -b(t)\sum_j \quad (-1)^j \hat{S}_j^z \qquad (39)$$

where $b(t) \equiv -\eta_z^s B(t)$. While the **second laser-spin coupling** is the coupling of the laser electric field **E(t)** to the spin-dependent electric polarization proportional to $\hat{S}_j \cdot \hat{S}_{j+1}$, which is called magnetostriction (MS) effect [20]. The MS coupling term is:

$$\hat{H}_{ext}^{MS}(t) = -E(t)\sum_j \ [\eta_{MS}^u + (-1)^j \eta_{MS}^s](\ \hat{S}_j^x \hat{S}_{j+1}^x + \ \hat{S}_j^y \hat{S}_{j+1}^y + \ \hat{S}_j^z \hat{S}_{j+1}^z) \quad (40)$$

We write the MS coupling term as $\hat{H}_{ext}^{MS}(t) = -E(t)\sum_j \ [\eta_{MS}^u + (-1)^j \eta_{MS}^s] \hat{S}_j^z \hat{S}_{j+1}^z$, by considering the Ising Hamiltonian **Eq. 1**. Here, $\eta_{MS}^u$ is a constant for converting spin dot product to polarization and in the typical multiferroic materials is smaller than $\eta_{MS}^s$ [39-41].

Then the simplified MS coupling becomes [20]:

$$\hat{H}_{ext}^{MS}(t) = -e(t)\sum_j \ [(-1)^j \eta_{MS}^s] \hat{S}_j^z \hat{S}_{j+1}^z; \quad (41)$$

where $e(t) \equiv -\eta_z^s E(t)$. The first observable quantity; electric polarization is given by [20]:

$$\hat{P} = -\eta_{MS}^s \sum_j \ (-1)^j \hat{S}_j^z \hat{S}_{j+1}^z \quad (42)$$

The time expectation value of electric polarization; $P(t) = <\hat{P}>_t$ acts as the source of electromagnetic radiation, which will be detected by experimental set up. The intensity of the radiation power at frequency ω is obtained by [20]:

$$I_p(\omega) = |\omega^2 \ P(\omega)|^2 \quad (43)$$

here P(ω) as the radiation power is the Fourier transform of P(t). In the following, we compute $I_p(\omega)$, which corresponds to HHG by exhibiting several peaks at integer multiples of the frequency.

### A. Fermionization

To compute the laser-matter coupling and electric polarization, we then fermionize the coupling term to make matrix representation in our basis. Two coupling terms Zeeman and MS are given by [20]:

$$\widehat{H}^Z_{ext} = b(t)\sigma_z \qquad (44)$$

$$\widehat{H}^{MS}_{ext} = e(t)\, cosk\sigma_z \qquad (45)$$

and the electric polarization reads:

$$P(k) = -\eta^S_{MS} cosk\sigma_z \qquad (46)$$

The spin current $\hat{I}_{spin}$ depends on the internal and external coupling term (i.e. J), and b(t) and e(t), where its matrix representation for the Zeeman coupling [20]:

$$\hat{I}_{spin}(k,t) =$$

$$\sum_k \ (J - e - b(t))\,[(F - \sin(ka)\,\sigma_z + \cos(ka)\,\sigma_x] + (J')\,[\sin(2ka)\,\sigma_z +$$

$$\cos(2ka)\,\sigma_x] \qquad (47)$$

### B. Laser pulse and time evolution

In the present study, the spin array system is in the ground state (the fully occupied lower energy band and the unoccupied upper bands) and then the laser electric or magnetic field is turned on. Recently, there have been demonstrations of strong field amplitudes (1MV/cm) at THz regimes for pulse lasers [42-44], where the generation of THz continuous waves is difficult. Therefore, we consider the pulse shape of the laser magnetic field b(t) as [20]:

$$b(t) = b_0 \cos(\Omega t) f(t); \qquad (48)$$

here, $b_0$ is the peak coupling energy, $\Omega$ is the central frequency and f(t) is the Gaussian envelope function, $f(t) = exp\,(-\Omega ln2(t^2/t^2_{FWHM}))$ [20], where $t_{FWHM}$ represents the full width at half maximum of the intensity $b^2(t)$ (the $t_{FWHM}/T$ with $T=2\pi/\Omega$ as the number of cycles of the pulse field. Time evolution of wave function of our system is computed by a quantum master equation of the Lindblad form [20,45]:

$$\frac{d}{dt}\rho(k,t) = -i[H(k,t),\rho(k,t)] + \gamma(L_k\rho(k,t)L_k^+ - \frac{1}{2}\{L_k^+ L_k, \rho(k,t)\}) \tag{49}$$

here ρ(k,t) is the density matrix (ψψ*) of the wave number k-subspace. In **Eq. 49** the Lindblad operator $L_k \equiv |\psi_g(k)><\psi_e(k)|$, represents the relaxation from the excited state $|\psi_e(k)>$ to the ground state $|\psi_g(k)>$, and γ shows its rate and corresponds to the lifetime. In our calculation, we set γ = 0.1J and independent of k for simplicity.

### C. Units and scales of physical quantities

Before discussing our results, we make remarks on the scales of physical quantities. In the following, we work in the units with J = 1 and represent all physical quantities including the photon energy, the lifetime of the magnetic excitation, and the magnetic and the electric fields in a dimensionless manner with the physical constants set to unity. The rules to recover the units depend on the value of J that we suppose.

**Table 1. Table of units for physical parameters depending on two choices of J = 10K and 50 K.**

| Energy, **J** | 10K | 50K |
|---|---|---|
| Photon energy, ℏΩ | 0.86 meV | 4.3 meV |
| Time, ℏ/J | 0.76 ps | 0.15 ps |
| Frequency, f = Ω/2π | 0.21 THz | 1.0 THz |
| Magnetic field, J/gμ$_B$ | 7.4 T | 37 T |
| Electric filed, E$_0$ = cB$_0$ | 22 MV/cm | 112 MV/cm |

### D. Electric polarization and spin current

In this section, we discuss the high-harmonic spectrum of the electric polarization as described by **Eq. 38**, **Eq. 39**, and **Eq. 40;** and apply dimensionless units as reported in **Table 1**. Time profile of electric polarization $\hat{P}(t)$ for the ac Zeeman coupling laser pulse with frequency $\Omega = 0.05, 0.5$ is plotted in **Fig. 12a** and corresponding power spectrum $I_p(\omega)$ is plotted in **Fig. 12b**. The behavior of $\hat{P}(t)$ exhibits the harmonic peaks of electric polarization. We systematically compute the time evolution of spin current $\hat{I}_{spin}$ at ac

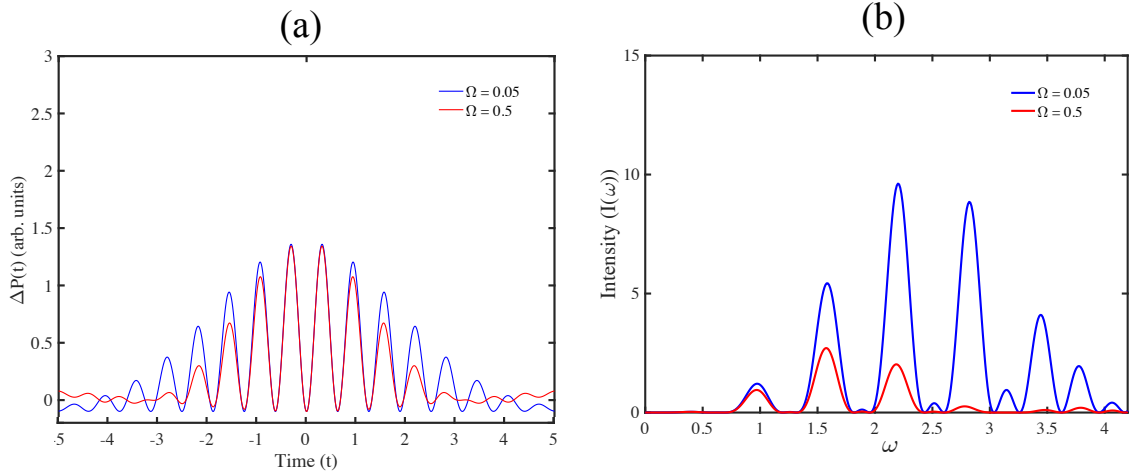

**Fig. 12:** (a) Time profile of polarization $\hat{P}(t)$ for the ac Zeeman with frequency $\Omega = 0.05, 0.5$. (b) Corresponding power spectrum.

Zeeman coupling with frequency $\Omega = 0.5$, which is computed by using **Eq. 46** and plotted in **Fig.13a** for unstrained H-SiB structure. This figure indicates the significant harmonic peaks. The intensity spectrum of spin current (**Eq. 47**) the second-harmonic (SHG) and third-harmonic generation (THG) is computed and plotted in **Fig. 13b** regards to unstrained H-SiB. In **Fig. 13b**, the intensity of the spin current of the SHG exhibits a nonmonotonic behavior as a function of J and $b_0$ (**Eq. 47**). The nonmonotonic behavior can be attributed to vanishing of the second harmonic in the limit of $b_0 \to 0$. The THG shows the same trend as SHG for unstrained structure.

We remark that **Fig. 14 and Fig. 15** indicate the intensity of SHG and THG spin current as a function of J and $b_0$ for strained structures of +6% and -6%, and for strained structure of +2%, +4% and -2%, -4% are brought in SI (Figs. S4-S7). It is worth noting that nonmonotonic behavior of intensity for second harmonic differs with third harmonic. As can be seen in these figures, different behaviors in the limit of $b_0 \to 0$ barises regard to strain value in line with **Eq. 47.** From **Fig. 13 to Fig. 15**, we observe that the strain and elastic motion cause the magnetic dynamic and shift the spin current. These findings suggest that spin current diffuse due to mechanical deformation where the structure subjected to the time dependent filed.

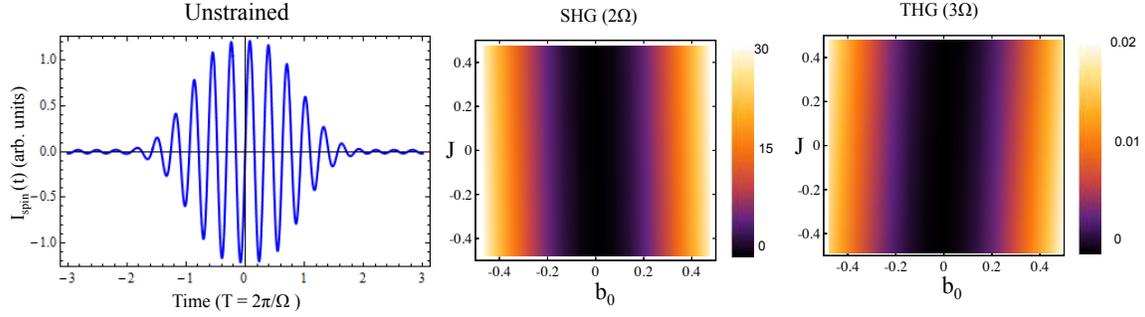

**Fig. 13:** Time evolution of spin current $\hat{I}_{spin}(t)$ for the ac Zeeman coupling with frequency $\Omega = 0.5$. Intensity of the spin currents of the second-harmonic generation (SHG) $[\hat{I}_{spin}(2\Omega)]$ (left) and third-harmonic generation (THG) spin currents $[\hat{I}_{spin}(3\Omega)]$ (right) by the ac Zeeman coupling for unstrained structure.

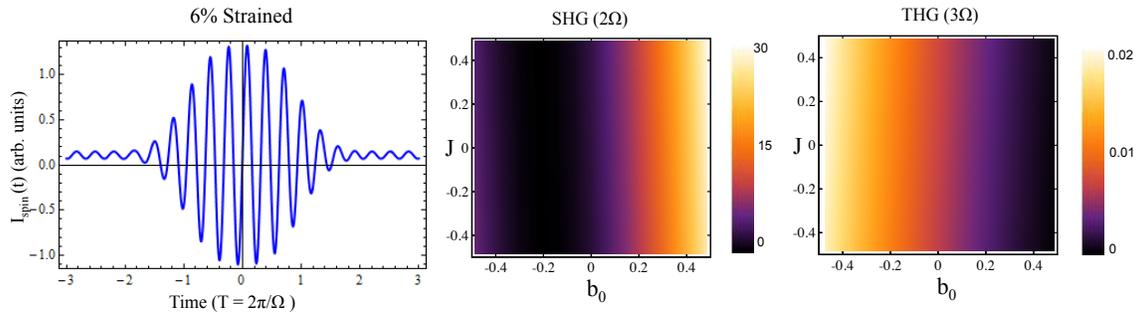

**Fig. 14:** Time evolution of spin current $\hat{I}_{spin}(t)$ for the ac Zeeman coupling with frequency $\Omega = 0.5$ for 6% strained H-SiB structure. Intensity of the spin currents of the second-harmonic generation (SHG) $[\hat{I}_{spin}(2\Omega)]$ (left) and third-harmonic generation (THG) spin currents $[\hat{I}_{spin}(3\Omega)]$ (right) by the ac Zeeman coupling for 6% strained structure.

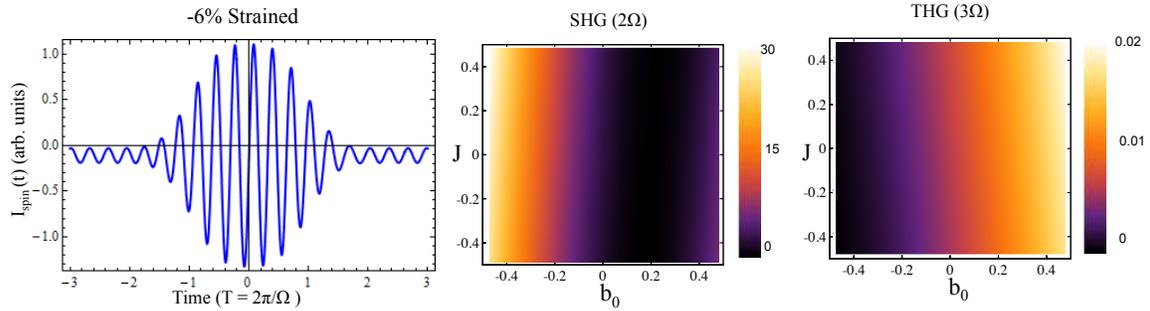

**Fig. 15:** Time evolution of spin current $\hat{I}_{spin}(t)$ for the ac Zeeman coupling with frequency $\Omega = 0.5$ for -4% and -6% strained H-SiB structure. Intensity of the spin currents of the second-harmonic generation (SHG) [$\hat{I}_{spin}(2\Omega)$] (left) and third-harmonic generation (THG) spin currents [$\hat{I}_{spin}(3\Omega)$] (right) by the ac Zeeman coupling for -6% strained H-SiB structure.

## VI. CONCLUSION

In summary, we acquire the phase diagram of 2D H-SiB semiconductors by means of $\pi$-rings Ising model and adiabatic quantum computing (AQC) and Monte Carlo simulation. We described a formation of disordered states arising at very low temperatures. To figure out the glassy, order and disorder states, the average of domain-wall lengths and peaks of heat capacity are calculated for unstrained and strained H-SiB structure. The origin of disorder states is the proliferation of topological defects. The disordered states evolve by decreasing temperature from ordered states due to increased relaxation time for nonequilibrium domain-wall lengths since the temperature decreases. Our findings reveal that the states in unstrained and strained 2D H-SiB are switched adiabatically with variation of coupling constant, which is applicable to AQC as a special type of memory elements. Moreover, the spin current and intensity of second and third harmonic generation (SHG) and (THG) have been calculated for magnetic insulator of H-SiB system by implementing of spin-laser coupling, i.e. the interaction of electromagnetic filed with spin planes of electric polarization of H-SiB. Spin planes has been mapped to two-band fermions and use quantum master equation for time evolution of the spin current, where HHG as a simple nonlinear optical effect computed for magnetic insulator of H-SiB. The present work can be mapped to investigate the special materials such as magnons with the Jordan-Wigner fermions, which have potential to resonate with an external field and create stronger signals for sophisticated experiments.

**Conflict of interest**

The authors declare no competing financial interest.

**References**

1. R. J. Elliott, Phys. Rev. 124, 346 (1961).
2. M. E. Fisher and W. Selke, Phys. Rev. Lett. 44, 1502 (1980).
3. S. Sachev, Quantum Phase Transitions, The Cambridge University Press, Cambridge, (1999).
4. P. Bak, Phys. Today 39, 38 (1986).
5. P. Bak and R. B. Bruinsma, Phys. Rev. Lett. 49, 249 (1982).
6. F. V. Kusmartsev, Phys. Rev. Lett. 69, 2268 (1992).
7. E. Farhi, J. Goldstone, S. Gutmann, and M. Sipser, arXiv:quantph/0001106.
8. V. Kusmartsev, D. M. Forrester, and M. S. Garelli, Napoli, (2005).
9. A. O'Hare, F. V. Kusmartsev, K. I. Kugel and M. S. Laad, *Phys. Rev. B*, 76, 064528 (2007).
10. F. Calegari, G. Sansone, S. Stagira, C. Vozzi, and M. Nisoli, Journal of Physics B: Atomic, Molecular and Optical Physics 49, 062001 (2016).
11. T. Brabec and F. Krausz, Reviews of Modern Physics 72, 545 (2000).
12. T. Brabec and F. Krausz, Reviews of Modern Physics 72, 545 (2000).
13. G. Ghimire, Ndabashimiye, A. D. DiChiara, E. Sistrunk, M. I. Stockman, P. Agostini, L. F. DiMauro, and D. A. Reis, Journal of Physics B: Atomic, Molecular and Optical Physics 47, 204030 (2014).
14. O. Schubert, M. Hohenleutner, F. Langer, B. Urbanek, C. Lange, U. Huttner, D. Golde, T. Meier, M. Kira, S. W. Koch, and R. Huber, Nature Photonics 8, 119 (2014).
15. M. Hohenleutner, F. Langer, O. Schubert, M. Knorr, U. Huttner, S. W. Koch, M. Kira, and R. Huber, Nature 523, 572 (2015).
16. T. T. Luu, M. Garg, S. Y. Kruchinin, A. Moulet, M. T. Hassan, and E. Goulielmakis, Nature 521, 498 (2015).
17. G. Ndabashimiye, S. Ghimire, M. Wu, D. A. Browne,K. J. Schafer, M. B. Gaarde, and D. A. Reis, Nature 534, 520 (2016).
18. G. Vampa, B. G. Ghamsari, S. Siadat Mousavi, T. J.Hammond, A. Olivieri, E. Lisicka-Skrek, A. Y. Naumov,D. M. Villeneuve, A. Staudte, P. Berini, and P. B. Corkum, Nature Physics 13, 659 (2017).
19. K. Kaneshima, Y. Shinohara, K. Takeuchi, N. Ishii,K. Imasaka, T. Kaji, S. Ashihara, K. L. Ishikawa, and J. Itatani, Physical Review Letters 120, 243903 (2018).
20. T. N. Ikeda, M. Sato, M. Phys. Rev. B 100, 214424 (2019).
21. H. Hirori, A. Doi, F. Blanchard, and K. Tanaka, Applied Physics Letters 98, 91106 (2011).

22. A. Ramazani, F. Shayeganfar, J. Jalilian, N. Fang, Nanophotonics, 9 (2), 337-349 (2020).
23. A. Hansson, F. de Brito Mota, R. Rivelino, Phys. Rev. B, 86, 195416 (2012).
24. Y. Ding, Y. Wang, J. Phys. Chem. C 117 (35), 18266-18278 (2013).
25. J. Dai, Y. Zhao, X. Wu, J. Yang, XC. Zeng, J. Phys. Chem. Lett 4 (4), 561-567 (2013).
26. A. O'Hare, F. V. Kusmartsev and K. I. Kugel, Nano Lett, 12, 1045−1052 (2012).
27. G.G. Naumis, S. Barraza-Lopez, M. Oliva-Leyva, H. Terrones, Rep. Prog. Phys. (80) 096501 (2017).
28. F. V. Kusmartsev, E. I. Rashba, EZh. Eksp. Teor. Fiz. 1984, 86, 142, Sov. Phys. JETP 1984, 59, 668.
29. A. O'Hare, F. V. Kusmartsev and K. I. Kugel, Phys. Rev. B, 79, 014439 (2009).
30. S. H. W. van der Ploeg, A. Izmalkov, Alec Maassen van den Brink, U. Hübner, M. Grajcar, E. Il'ichev, H.-G. Meyer, and A. M. Zagoskin, Phys. Rev. Lett. 98, 057004 (2007).
31. S. H. W. van der Ploeg, A. Izmalkov, M. Grajcar, U. Hübner, S. Linzen, S. Uchaikin, Th. Wagner, A. Yu. Smirnov, A. Maassen van den Brink, M. H. S. Amin, A. M. Zagoskin, E. Il'ichev, and H.-G. Meyer, arXiv:cond-mat/0702580
32. S. H. W. van der Ploeg, A. Izmalkov, Alec Maassen van den Brink, U. Hübner, M. Grajcar, E. Il'ichev, H.-G. Meyer, and A. M. Zagoskin, Phys. Rev. Lett., 98, 057004 (2007).
33. H. Ishizuka and M. Sato, Phys. Rev. Lett. 122, 197702.
34. H. Ishizuka and M. Sato, (2019), arXiv:1907.02734.
35. S. Takayoshi, H. Aoki, and T. Oka, Physical Review B 90, 085150 (2014), arXiv:1302.4460.
36. S. Takayoshi, M. Sato, and T. Oka, Physical Review B 90, 214413 (2014), arXiv:1402.0881.
37. M. Sato, S. Takayoshi, and T. Oka, Physical Review Letters 117, 147202 (2016), arXiv:1602.03702.
38. M. Sato, Y. Sasaki, and T. Oka, (2014), arXiv:1404.2010.
39. A. Pimenov, A. A. Mukhin, V. Y. Ivanov, V. D. Travkin, A. M. Balbashov, and A. Loidl, Nature Physics 2, 97 (2006).
40. Y. Takahashi, R. Shimano, Y. Kaneko, H. Murakawa, and Y. Tokura, Nature Physics 8, 121 (2011).
41. T. Kubacka, J. A. Johnson, M. C. Hoffmann, C. Vicario, S. de Jong, P. Beaud, S. Grubel, S.-W. Huang, L. Huber, et al., Science 343, 1333 (2014).
42. H. Hirori, A. Doi, F. Blanchard, and K. Tanaka, Applied Physics Letters 98, 91106 (2011).
43. B. Liu, H. Bromberger, A. Cartella, T. Gebert, M. Först, and A. Cavalleri, Opt. Lett. 42, 129 (2017).
44. Y. Mukai, H. Hirori, T. Yamamoto, H. Kageyama, and K. Tanaka, New Journal of Physics 18, 13045 (2016).
45. S. Sachdev, Quantum Phase Transitions (Cambridge University Press, 2011).

# Supporting Information
# Terahertz Nonlinear Optics in Two-dimensional Semi-Hydrogenated SiB

Ali Ramazani[1], Farzaneh Shayeganfar[2,3], Anshuman Kumar[4], Nicholas X Fang[1, †]

[1]Department of Mechanical Engineering, Massachusetts Institute of Technology, Cambridge, MA 02139, USA
[2]Department of Civil and Environmental Engineering, Rice University, Houston, TX 77005, USA
[3]Department of Physics and Energy Engineering, Amirkabir University of Technology
[4]Physics Department, Indian Institute of Technology Bombay, Mumbai 400076, India

**S1. Topological defects (Defected boundary)**

In this section, we discuss the possible types of defects (such as dislocation) and phase boundaries, which could arise in our model. We start with analyzing the most natural defect for this model, which is the boundary between the domains corresponding to double and single stripe (**Fig. S1**). We calculate the energy (per site) for the flat portion (fp) of this domain boundary. We take a six-site plaquette around point A and calculate energies of sites numbered from 1 to 6. Thus, we suppose that the atom 1 is H-Si site and we acquire:

---

* Corresponding Author; E-mail: nicfang@mit.edu

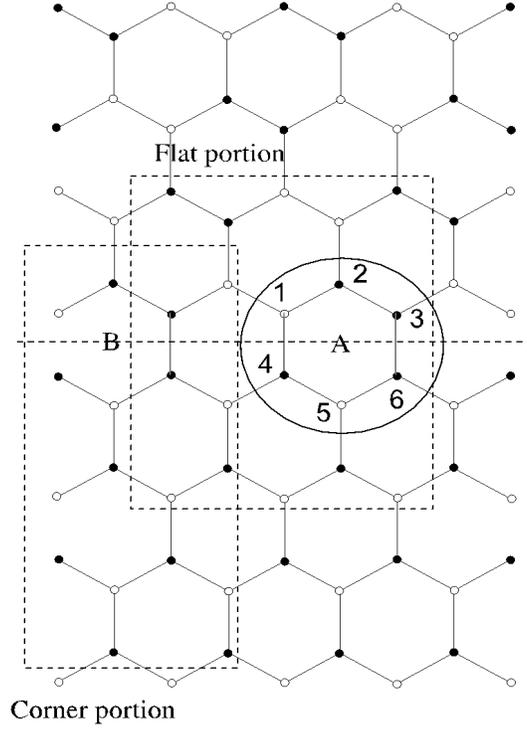

**Fig. S1:** Flat and corner portion of double and single stripe of hexagonal structures.

$$E_1 = \tfrac{2}{3}(-4J + 2J + 12J' - 12J') = -\tfrac{4}{3}J \ , \ E_2 = \tfrac{2}{3}(-4J + 2J + 3J' - 3J') = -\tfrac{4}{3}J$$

$$E_3 = \tfrac{2}{3}(2J - 16J') = \tfrac{4}{3}J - \tfrac{32}{3}J' \ , \quad E_4 = \tfrac{2}{3}(6J + 4J') = 4J + \tfrac{8}{3}J',$$

$$E_5 = \tfrac{2}{3}(-6J + 16J') = -4J + \tfrac{32}{3}J', \quad E_6 = \tfrac{2}{3}(-2J + 4J') = -\tfrac{4}{3}J + \tfrac{8}{3}J'$$

$$E_{fp} = -\tfrac{8}{3}J + \tfrac{16}{3}J' \tag{S1}$$

The same calculation for point B in **Fig. S1** proves that $E_{cp}$ is equal to $E_{fp}$, that is:

$$E_{fp} = -\tfrac{8}{3}J + \tfrac{16}{3}J' \tag{S2}$$

This result suggests that the formation of domain boundaries at finite temperatures could be favorable from the viewpoint of the configuration entropy. In the next step, we consider to a dislocation (shift by one lattice constant) in the phases with single and double stripes as shown in **Fig. S2.**

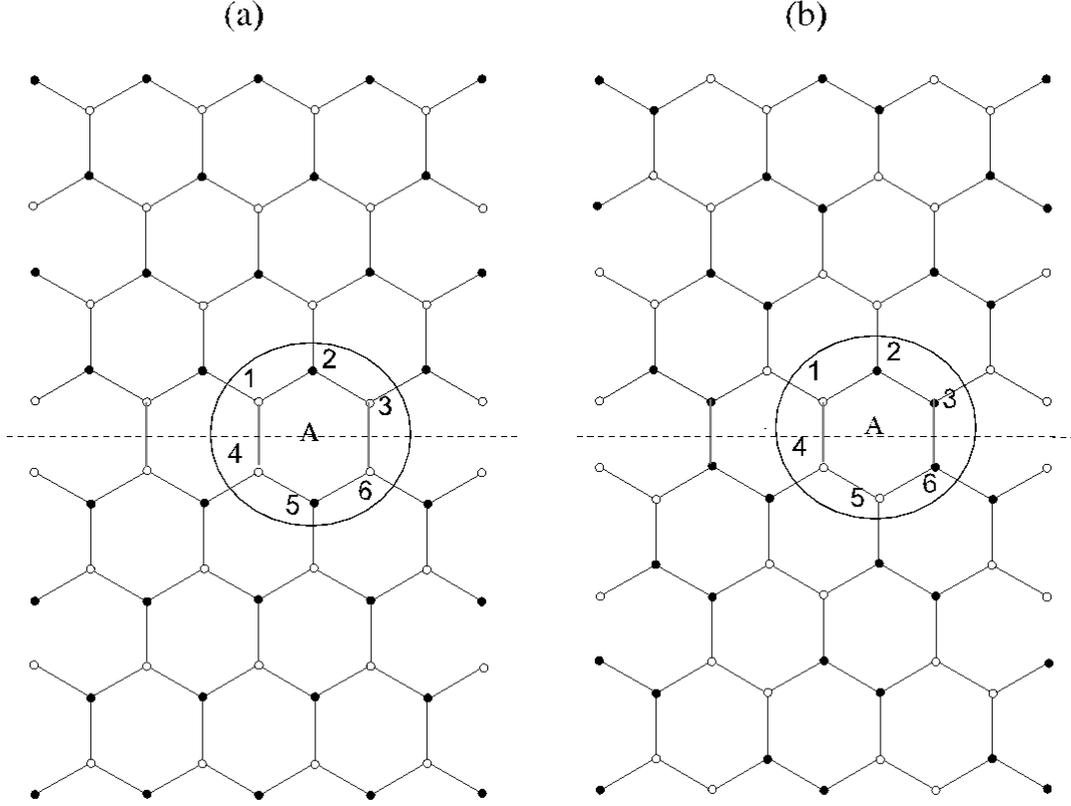

**Fig. S2:** A dislocation (shift by one lattice constant) in the phases with single (a) and double (b) stripes.

We can compute the dislocation portion energy ($E_{dp}$) for the left panel in **Fig. S2a**:

$E_1 = \frac{2}{3}(-2J + 8J') = -\frac{4}{3}J + \frac{16}{3}J'$ ; $E_2 = \frac{2}{3}(-6J + 2J') = -4J + \frac{4}{3}J'$

$E_3 = \frac{2}{3}(-2J + 8J') = -\frac{4}{3}J + \frac{16}{3}J'$ ; $E_4 = \frac{2}{3}(-2J + 2J') = -\frac{4}{3}J + \frac{4}{3}J'$

$E_5 = \frac{2}{3}(-6J + 8J') = -4J + \frac{16}{3}J'$ ; $E_6 = \frac{2}{3}(-2J + 2J') = -\frac{4}{3}J + \frac{4}{3}J'$

$$E_{dp} = -\frac{40}{3}J + 20J' \tag{S3}$$

and the same calculation for $E_{dp}$ of right panel in **Fig. S2b**:

$$E_1 = \tfrac{2}{3}(+2J - 8J') = +\tfrac{4}{3}J - \tfrac{16}{3}J' \quad ; \quad E_2 = \tfrac{2}{3}(-2J - 2J') = -\tfrac{4}{3}J - \tfrac{4}{3}J'$$

$$E_3 = \tfrac{2}{3}(+2J - 8J') = +\tfrac{4}{3}J - \tfrac{16}{3}J' \quad ; \quad E_4 = \tfrac{2}{3}(+2J - 2J') = +\tfrac{4}{3}J - \tfrac{4}{3}J'$$

$$E_5 = \tfrac{2}{3}(-2J - 8J') = -\tfrac{4}{3}J - \tfrac{16}{3}J' \quad ; \quad E_6 = \tfrac{2}{3}(+2J - 2J') = +\tfrac{4}{3}J - \tfrac{4}{3}J'$$

$$E_{dp} = +\tfrac{8}{3}J - 20J' \tag{S4}$$

These results mean that the formation of dislocation for **Fig. S2a** (single stripe) at finite temperatures could be more favorable than dislocation of **Fig. S2b** (double stripe) from the viewpoint of the configuration entropy.

To summarize, the entropy of a state with the boundary of single stripe is significantly larger than the entropy of a double stripe dislocation. This implies the possibility of the creation of single dislocations near the crossover between these antiferromagnetic and ordered phases. Therefore, at nonzero low temperatures, the boundary and single stripe dislocation defects described above will proliferate into the ordered antiferromagnetic states.

## S2. Energy values of strained Hamiltonian

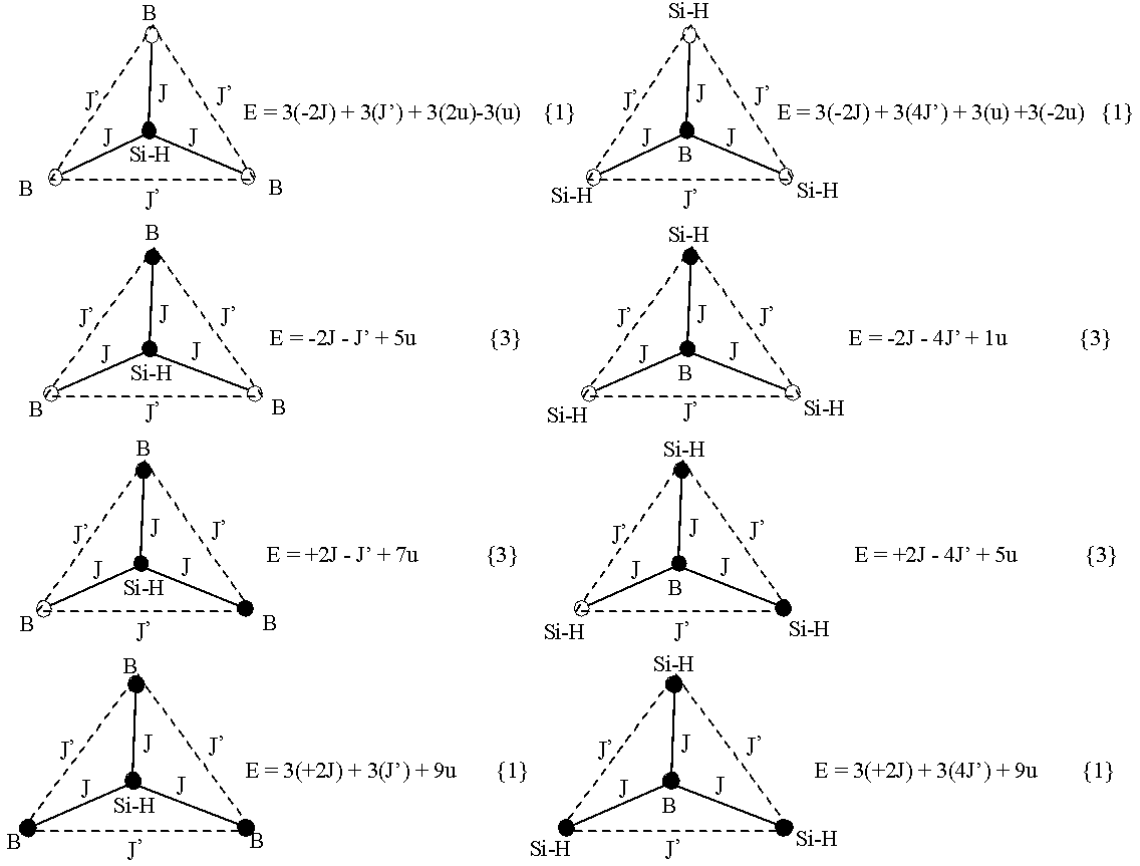

**Fig. S3.** Similar to **Fig. 2**, energy value of possible spin configurations of a four-site plaquette by considering extended Hamiltonian **Eq. 22,** which considers spin-lattice; the degree of degeneracy of each energy value is shown in the brackets. Here, only 16 configurations with the filled circle in the center are shown; where there are 16 similar configurations with the open circle in the center of the plaquette.

## S3. The Bogoliubov transformation

To gain a better insight to the Ising Hamiltonian of **Eq. 34** in the main text, we introduce the following Fourier transformations and Bogoliubov transforming to get exact solution. We start with the fermionic annihilation operator $c_j$, where $c_k$ is its Fourier transformed.

To ensure the reality of the Hamiltonian, we introduce a phase factor in the Fourier transformation as below[1]:

$$c_k = \frac{e^{-i\pi/4}}{\sqrt{N}} \sum_j c_j e^{-ikx_j} \qquad (S5)$$

after substituting in **Eq. 34** in the main text, the Ising Hamiltonian becomes:

$$H = \sum_k 2(JF - J\cos(ka))\hat{c}_k^+ \hat{c}_k + J\sin(ka)(\hat{c}_{-k}^+\hat{c}_k^+ - \hat{c}_{-k}\hat{c}_k) - JF \qquad (S6)$$

where, $a$ is the distance between two sites in the hexagonal Ising model, N is the number sites in the structure model, k is the wave number ($k = 2\pi n/Na$), and $x_i = ia$, the position of the i-th site. The number n by considering periodic boundary takes the values **[S1]**:

$$n = -\frac{N-1}{2}, \ldots, \frac{N-1}{2} \qquad \text{for n odd} \qquad (S7)$$

$$n = -\frac{N}{2}, \ldots, \frac{N}{2} \qquad \text{for n even} \qquad (S8)$$

In the next step, we diagonalize the Hamiltonian (**Eq. S6**) by applying Bogoliubov transformation. For this transformation, we define $\gamma_k$ and $\gamma_k^+$ by $u_k$ and $v_k$ coefficients, which satisfy $u_k^2 + v_k^2 = 1$, $u_{-k} = u_k$, and $v_{-k} = v_k$,

$$\gamma_k = u_k c_k - v_k c_{-k}^+$$

(S9)

$$\gamma_k^+ = u_k c_k^+ - v_k c_{-k}$$

(S10)

As it turns out, the following choice for $u_k$ and $v_k$ suffices **[S1]**:

$$u_k = \cos\left(\frac{\theta_k}{2}\right), \quad v_k = \sin\left(\frac{\theta_k}{2}\right)$$

(S11)

$$\tan(\theta_k) = \frac{\sin(ka)}{F - \cos(ka)}$$

(S12)

to fulfill the above requirements. Moreover, the commutation relations are preserved by this transformation **[S1]**:

$$\{\gamma_k, \gamma_l^+\} = \delta_{kl} \ , \ \{\gamma_k^+, \gamma_l^+\} = \{\gamma_k, \gamma_l\} = 0$$

(S13)

and the Hamiltonian turns out to be diagonalized by this choice of basis **[S1]**:

$$H_I = \sum_K \epsilon_k \left(\gamma_k^+ \gamma_k - \frac{1}{2}\right)$$

(S14)

$$\epsilon_K = 2J \sqrt{1 + F^2 - 2F \cos(ka)}$$

(S15)

The spin current as defined in **Eq. 25** in the main text depends on the coupling term, and its matrix representation is:

$$H = 2J \sum_k \ (F - \cos(ka))\sigma_z + \sin(ka)\sigma_y \tag{S16}$$

**S4. Spin current and intensities of SHG for strained H-SiB**

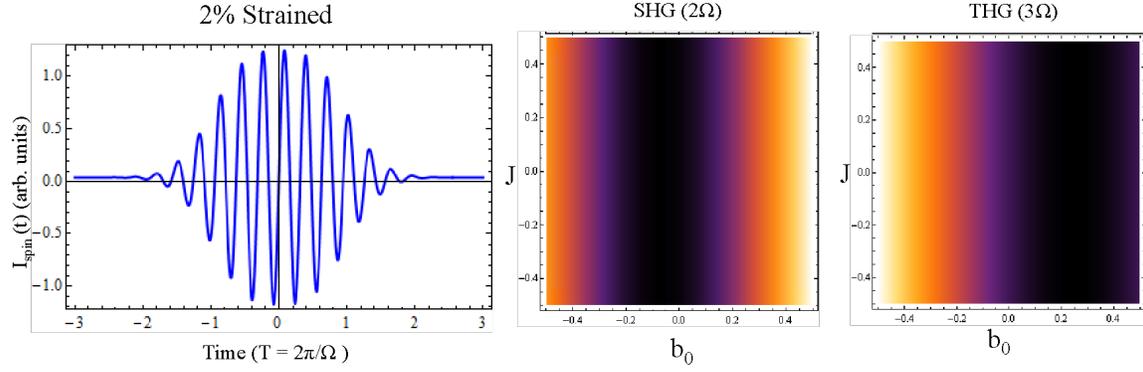

**Fig. S4:** Time evolution of spin current $\hat{I}_{spin}(t)$ for the ac Zeeman coupling with frequency $\Omega = 0.5$ for 2% strained H-SiB structure. Intensities of second-harmonic generation (SHG) $[\hat{I}_{spin}(2\Omega)]$ (left) and third-harmonic generation (THG) spin currents $[\hat{I}_{spin}(3\Omega)]$ (right) by the ac Zeeman coupling for 2% strained structure.

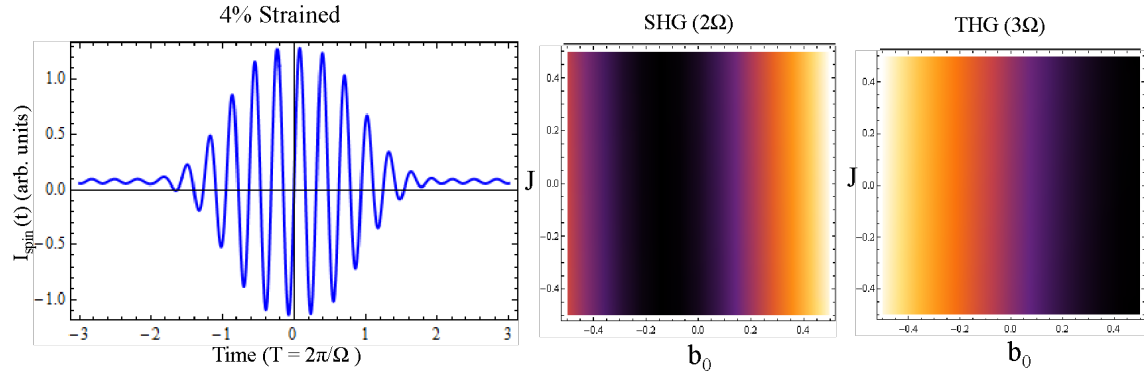

**Fig. S5:** Time evolution of spin current $\hat{I}_{spin}(t)$ for the ac Zeeman coupling with frequency $\Omega = 0.5$ for 4% strained H-SiB structure. Intensities of second-harmonic generation (SHG) $[\hat{I}_{spin}(2\Omega)]$ (left) and third-harmonic generation (THG) spin currents $[\hat{I}_{spin}(3\Omega)]$ (right) by the ac Zeeman coupling for 4% strained structure.

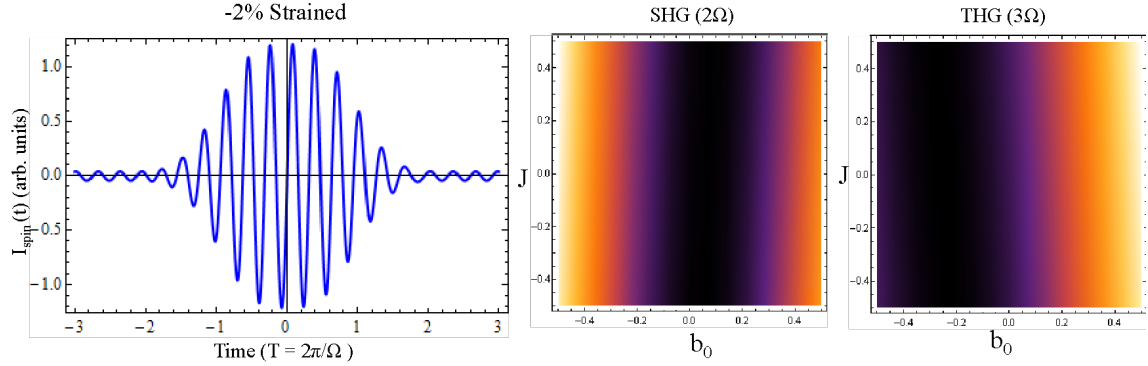

**Fig. S6:** Time evolution of spin current $\hat{I}_{spin}(t)$ for the ac Zeeman coupling with frequency $\Omega = 0.5$ for -2% strained H-SiB structure. Intensities of second-harmonic generation (SHG) [$\hat{I}_{spin}(2\Omega)$] (left) and third-harmonic generation (THG) spin currents [$\hat{I}_{spin}(3\Omega)$] (right) by the ac Zeeman coupling for -2% strained structure.

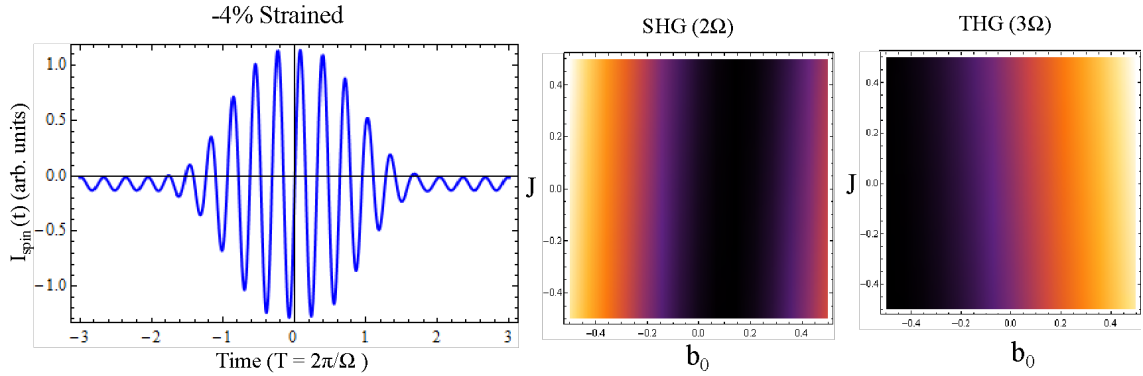

**Fig. S7:** Time evolution of spin current $\hat{I}_{spin}(t)$ for the ac Zeeman coupling with frequency $\Omega = 0.5$ for -4% and -6% strained H-SiB structure. Intensities of second-harmonic generation (SHG) [$\hat{I}_{spin}(2\Omega)$] (left) and third-harmonic generation (THG) spin currents [$\hat{I}_{spin}(3\Omega)$] (right) by the ac Zeeman coupling for -4% strained H-SiB structure.

**References:**

S1. S. H. W. van der Ploeg, A. Izmalkov, Alec Maassen van den Brink, U. Hübner, M. Grajcar, E. Il'ichev, H.-G. Meyer, and A. M. Zagoskin, *Phys. Rev. Lett.*, 98, 057004 (2007).